\documentclass[12pt]{article}
\usepackage{latexsym}
\usepackage{mathrsfs}
\usepackage{amssymb}
\usepackage{amsmath}
\usepackage{amsfonts, epsf,epsfig,color}
\usepackage{graphicx}
\usepackage{tensor}
\usepackage[all]{xy}
\include{macro}

\newcommand{\C}{\mathbb{C}}

\newcommand{\R}{\mathbb{R}}
\renewcommand{\P}{\mathbb{P}}

\newcommand{\F}{\mathbb{F}}

\newcommand{\M}{\mathbb{M}}

\newcommand{\T}{\mathbb{T}}
\newcommand{\Z}{\mathbb{Z}}
\newcommand{\p}{\partial}

\newcommand{\D}{\mathrm{D}}

\renewcommand{\P}{\mathbb{P}}

\newcommand{\rd}{\, \mathrm{d}}

\newcommand{\be}{\begin{equation}\label}
\newcommand{\ee}{\end{equation}}
\newcommand{\bea}{\begin{eqnarray}\label}
\newcommand{\eea}{\end{eqnarray}}

\begin{document}

\title{\textbf{On Closed Twistor String Theory}}

\author{R A Reid-Edwards\footnote{reidedwards@maths.ox.ac.uk}\\
\\
The Mathematical Institute\\
University of Oxford\\
24-29 St.~Giles', Oxford OX1 3LB\\
United Kingdom\\
}

\date{}

\maketitle

\vspace{3cm}

\begin{abstract}
The Heterotic twistor string theory of Mason and Skinner is investigated with particular attention given to the role of topological gravity on the world-sheet. The general structure of scattering amplitudes is discussed and expressed in terms of an integral over the moduli space of super Riemann surfaces. 
\end{abstract}

\newpage

\section{Introduction}

Twistor string theory introduced by Witten in \cite{Witten:2003nn} is thought to be equivalent, via the Penrose transform, to an exotic version of ${\cal N}=4$ conformal supergravity in space-time \cite{Berkovits:2004jj}. Though the presence of conformal, rather than Einstein, gravity makes the theory less attractive, there are still many interesting features that make it worthy of further investigation and recent progress has lead to a better understanding of the theory \cite{Dolan:2007vv,Dolan:2008gc,Adamo:2012xe,Adamo:2012nn}. The framework introduced by twistor string theory has lead to new ways of introducing non-linear interactions in twistor space \cite{Mason:2005zm,Boels:2007qn,Boels:2006ir,Adamo:2011cb}, scattering amplitudes and recursion relations for the ${\cal N}=4$ Yang-Mills sector of the theory take on a much simpler and manifestly symmetric form in twistor space \cite{ArkaniHamed:2009dn,ArkaniHamed:2010kv}. The equivalence between Wilson loops and scattering amplitudes becomes much clearer when considered using momentum twistors (see for example \cite{Adamo:2011pv} and references contained therein). Many of these advances, though taking inspiration from twistor string theory, have developed in a way that is logically-independent of twistor string theory. The shortcomings of the twistor string in no way diminish or undermine the remarkable progress made in understanding perturbative ${\cal N}=4$ Yang-Mills in twistor (and momentum twistor) space. So why return to the original twistor string construction? The first reason is that there is unfinished business: Despite its shortcomings, the twistor string is an unusual theory in which a number of features of conventional and topological string theories appear in a novel way, many basic aspects of which are still poorly understood; The exact form of the rather exotic form of conformal supergravity theory relevant to twistor string theory \cite{Berkovits:2004jj} is still not properly understood and the rank of the gauge group required for the theory to be conformal \cite{Witten:2003nn,Dolan:2007vv} on the world-sheet has not been independently understood from the space-time perspective; The best studied twistor string theory is an open theory on split-signature space-time and it would be good to have a better understanding of the closed twistor string theories appropriate to other space-time signatures.

In addition to the above, there are more forward-looking reasons to return to the twistor string. Elegant descriptions of tree-level scattering amplitudes for ${\cal N}=8$ Einstein supergravity \cite{Cachazo:2012kg}, following the work of \cite{Hodges:2012ym,Hodges:2011wm}, were formulated in terms of rational maps into twistor space and have raised again the question of whether there is a twistor string description of ${\cal N}=8$ Einstein supergravity. The utility of such a description might lie in its ability to determine whether or not the theory is perturbatively finite. Twistor string theory also provides a rare example of a string theory for which we have a very good knowledge of the perturbative scattering amplitudes and may shed some light on alternative ways to think about conventional superstring perturbation theory.

An alternative description of Witten's theory was given by Berkovits in \cite{Berkovits:2004hg}, which is an open string theory with ends on a real slice of twistor space. A closed heterotic counterpart was proposed in \cite{Mason:2007zv} and various constructions of self-dual gravity and self-dual Yang-Mills are known \cite{AbouZeid:2006wu}. The Heterotic formulation of the closed twistor string is probably the one closest to conventional string theories and therefore the one which should be the simplest to understand in detail; however, a number of outstanding issues still remain to be clarified. Among them is the role of world-sheet gravity in the Heterotic twistor string. In sections two and three of this article, it is argued that the correct theory of world-sheet gravity is a `half-twsited' version of the two-dimensional topological gravity introduced in \cite{Witten:1988xj}. This is in contrast to the conventional world-sheet gravity used to describe the twistor string theory of \cite{Berkovits:2004hg}. Towards the end of section two, the `half-twisted' or (2,0) sigma model is reviewed. This sigma model, studied at length at the perturbative level in \cite{Witten:2005px}, forms the basis of the Heterotic twistor string theory \cite{Mason:2007zv}. In section four the path integral description of scattering amplitudes is discussed. The topological nature of the world-sheet gravity allows scattering amplitudes to be written in terms of an integral over the moduli space of super-Riemann surfaces in a way that is reminiscent of similar treatments of the conventional superstring. Spurred on by the apparent naturalness of topological, rather than conventional, world-sheet gravity, in the twistor string, it is suggested in section five that it might also be more appropriate to have topological gravity in the open twistor string. Crucially, gauge-fixing topological gravity requires additional ghost fields to be introduced to gauge fix the model and so changes the field content of the open string theory. The effect of this is to reduce the rank of the gauge group of the Yang-Mills sector. The salient features of twistor theory are briefly reviewed in Appendix A and Appendix B gives more details of an $n$-point gluon scattering amplitude calculation sketched in section five.

\section{Topological Gravity in Two Dimensions}

One of original motivations for studying topological gravity was the hope that it might lead to a more fundamental description of space-time from which conventional Einstein gravity might arise as a broken phase \cite{Witten:1988xi,Witten:1989ig}. Furthermore, in the context of two-dimensional sigma models, a topological sector alters the conformal anomaly in such a way as to ensure the theory is conformal in any dimension. The corresponding idea for string theory would then be that the critical dimension requirement of conventional string theory might arise from a topological theory by some symmetry breaking mechanism. Despite these lofty aims, topological field theory has really found its home in mathematical physics in the calculation of topological invariants such as intersection numbers \cite{Witten:1988ze,Hori:2003ic} where the gravitational aspect has often been of only secondary interest. Similarly, the potential role of topological gravity in twistor string theory has not been the subject of much discussion.

In this section two-dimensional topological gravity is reviewed. The treatment will closely follow \cite{Witten:1988xj}, following on from the four-dimensional construction presented in \cite{Witten:1988xi} and further details may be found in these papers (see also \cite{Witten:1989ig,Labastida:1988zb,Verlinde:1990ku} and the review \cite{Dijkgraaf:1990qw}). A slight difference here will be to treat the symmetries at the coordinate, rather then frame level (as done in \cite{Witten:1988xj}) but the relationship between the two descriptions is straightforward. The construction of \cite{Witten:1988xj} introduces a fermionic scalar operator $Q$ for which $Q^2$ generates an action of local diffeomorphisms and Weyl transformations in two dimensions.  A variation of this construction that is more suitable for application to the (2,0) theory, which forms the basis of the Heterotic twistor string, and in which the generator $Q$ is now nilpotent, will be discussed. This latter version of topological gravity admits a truncation to a (2,0) topological gravity which can be coupled to a (2,0) sigma model to give the `half-twisted' string theory which is the focus of this article.

\subsection{Pure Two-Dimensional Topological Gravity}

In order to fix notation, the world-sheet $\Sigma$ has coordinates $\sigma^{\alpha}$, where $\alpha=1,2$. It will often be useful to use light-cone $(\sigma^-,\sigma^+)$ or complex $(\sigma,\bar{\sigma})$ coordinates. Frame indices, when used, shall be denoted by $a=1,2$ (light-cone notation for frame indices will also be frequently used). The context should hopefully avoid any confusion between frame and coordinate indices that might arise. The frame and coordinate valued quantities are related by the zweibein $e_{\alpha}{}^a$ and its inverse $e_a{}^{\alpha}$ and the frame metric $\eta_{ab}$ is related to the world-sheet metric $h_{\alpha\beta}$ by 
$$
h_{\alpha\beta}=\eta_{ab}e^a_{\alpha}e^b_{\beta}\;,
$$
where $\eta_{ab}$ will usually be taken to be the invariant of $SO(1,1)$, except in section five, where it will be more useful to take it to be the invariant of $SO(2)$.

Following \cite{Witten:1988xj} a fermionic symmetry, for which the square of the generator gives an infinitesimal diffeomorphisms and Weyl transformations, is introduced. For example, the world-sheet field $\Phi$ transforms as
\begin{equation}\label{Qsquared}
Q^2\Phi={\cal L}_{\xi}\Phi+\delta_{\Omega}\Phi\;,
\end{equation}
where ${\cal L}_{\xi}\Phi$ is a Lie derivative with vector parameter $\xi=\xi^{\alpha}\partial_{\alpha}$ and $\delta_{\Omega}\Phi$ is a Weyl transformation with scalar parameter ${\Omega}$. A symmetric tensor $\chi_{\alpha\beta}$ and a vector $C_{\alpha}$, which transform under the $Q$, are introduced. $\chi_{\alpha\beta}$ is of spin two but, unlike the metric, is fermionic, and $C^{\alpha}$ is a bosonic world-sheet vector. The following infinitesimal action of the fermionic generator $Q$ on these fields is
\begin{eqnarray}
\delta_{\epsilon}h_{\alpha\beta}&=&2i\epsilon \chi_{\alpha\beta}\;,\nonumber\\
\delta_{\epsilon}\chi_{\alpha\beta}&=&\frac{1}{2}\epsilon \left(\nabla_{\alpha}C_{\beta}+\nabla_{\beta}C_{\alpha}- h_{\alpha\beta}\nabla_{\lambda}C^{\lambda}\right)\;,\nonumber\\
\delta_{\epsilon}C_{\alpha}&=&2i\epsilon\chi_{\alpha\beta}C^{\beta}\;,\nonumber\\
\delta_{\epsilon}C^{\alpha}&=&0\;.\nonumber
\end{eqnarray}
The fourth transformation follows from the first and the third and the fact that $C_{\alpha}=h_{\alpha\beta}C^{\beta}$. The transformations close to give an algebra in which the commutators act as two-dimensional conformal diffeomorphisms
\begin{eqnarray}
[\delta_{\epsilon_1},\delta_{\epsilon_2}]h_{\alpha\beta}&=& -2i\epsilon_1\epsilon_2\left(h_{\alpha\lambda}\nabla_{\beta}C^{\lambda}+h_{\beta\lambda}\nabla_{\alpha}C^{\lambda}-(\nabla_{\lambda}C^{\lambda})h_{\alpha\beta}\right)\;, \nonumber\\
\left[\delta_{\epsilon_1},\delta_{\epsilon_2}\right]\chi_{\alpha\beta}&=& -2i\epsilon_1\epsilon_2\left(C^{\lambda}\nabla_{\lambda}\chi_{\alpha\beta}+\chi_{\alpha\lambda}\nabla_{\beta}C^{\lambda}+\chi_{\beta\lambda}\nabla_{\alpha}C^{\lambda}-(\nabla_{\lambda}C^{\lambda})\chi_{\alpha\beta}\right)\;,\nonumber\\
\left[\delta_{\epsilon_1},\delta_{\epsilon_2}\right]C_{\alpha}&=&-2i\epsilon_1\epsilon_2\left(C^{\lambda}\nabla_{\alpha}C_{\lambda}+C^{\lambda}\nabla_{\lambda}C^{\alpha}-(\nabla_{\lambda}C^{\lambda})C_{\alpha}\right)\;.\nonumber
\end{eqnarray}
The right hand side of these expressions are simply conformal diffeomorphisms with bosonic parameters $\xi^{\lambda}=-2i\epsilon_1\epsilon_2C^{\lambda}$ and $\Omega=-2i\epsilon_1\epsilon_2\nabla_{\lambda}C^{\lambda}$.
One can try to construct kinetic terms for topological gravity with varying degrees of success \cite{Witten:1988xj,Labastida:1988zb}, but such attempts will not be of concern here.

\subsection{Sigma Models with Topological Gravity}

Topological sigma models give a natural candidate for a matter system to couple to two-dimensional conformal gravity. To fix notation, consider a sigma model on a world-sheet $\Sigma$ with embedding $X:\Sigma\rightarrow M$ into a $D$-dimensional target space $M$. In practice it will be assumed that $M$ is K\"{a}hler. In particular $M$ will be $\C\P^3$ for the twistor string; however, non-K\"{a}hler target spaces have been considered in \cite{Witten:1988xj}. The sigma model of interest has the scalar fields $X^I(\sigma)$ already mentioned,  one-forms $\rho^I(\sigma)=\rho^I_{\alpha}(\sigma)\rd \sigma^{\alpha}$, and world-sheet scalars $\alpha^I(\sigma)$ where $I=1,2,...D$. In contrast to conventional sigma models, the $\rho^I$ and $\alpha^I$ are fermionic (i.e. their spins are `twisted' with respect to what would be expected from the spin-statistics theorem \cite{Witten:1991zz}). It is also necessary to require that $\rho^I$ is self-dual $\rho^I=*J^I{}_J\rho^J$, where $J^I{}_J$ is the complex structure of the target space, and so $\rho^I$ represents only one world-sheet degree of freedom per target space dimension. To ensure the $Q$ symmetry closes off-shell, it is necessary to also include non-dynamical fields $H^I_{\alpha}$. Under the $Q$ symmetry the matter fields transform as \cite{Witten:1988xj}
\begin{eqnarray}
\delta_{\epsilon}X^I&=&i\epsilon \alpha^I\;,\nonumber\\
\delta_{\epsilon}\alpha^I&=&\epsilon C^{\alpha}\partial_{\alpha}X^I\;,\nonumber\\
\delta_{\epsilon}\rho^I_{\alpha}&=&\epsilon H^I_{\alpha}-i\epsilon\Gamma^I_{JK}\alpha^J\rho^K_{\alpha}+i\epsilon h_{\alpha\beta}\chi^{\beta\lambda}\rho^I_{\lambda}\;,\nonumber\\
\delta_{\epsilon}H^I_{\alpha}&=&-i\epsilon\Gamma^I_{JK}\alpha^JH^K_{\alpha}-\frac{1}{2}\epsilon R_{JKL}^I\alpha^J\alpha^K\rho^L_a+i\epsilon \left(C^{\beta}\nabla_{\beta}\rho^I_{\alpha}+\frac{i}{2}\varepsilon_{\alpha\beta}\Lambda\rho^{\beta I}+\frac{1}{2}(\nabla_{\beta}C^{\beta})\rho^I_{\alpha}\right)\;,\nonumber
\end{eqnarray}
where $\Lambda=-\varepsilon^{\rho\sigma}(\chi_{\rho\gamma}\chi^{\gamma}{}_{\sigma}-i\nabla_{\rho}C_{\sigma})$ and which close to give an algebra of the required form (\ref{Qsquared})
\begin{eqnarray}
[\delta_{\epsilon_1},\delta_{\epsilon_2}]X^I&=& -2i\epsilon_1\epsilon_2C^{\alpha}\partial_{\alpha}X^I\;, \nonumber\\
\left[\delta_{\epsilon_1},\delta_{\epsilon_2}\right]\alpha^I&=& -2i\epsilon_1\epsilon_2C^{\alpha}\partial_{\alpha}\alpha^I \;,\nonumber\\
\left[\delta_{\epsilon_1},\delta_{\epsilon_2}\right]\rho^I_{\alpha}&=&-2i\epsilon_1\epsilon_2\left(\rho^I_{\lambda}\p_{\alpha}C^{\lambda}+C^{\lambda}\p_{\lambda}\rho^I_{\alpha}\right)\;,\nonumber\\
\left[\delta_{\epsilon_1},\delta_{\epsilon_2}\right]H^I_{\alpha}&=&-2i\epsilon_1\epsilon_2\left(H^I_{\lambda}\p_{\alpha}C^{\lambda}+C^{\lambda}\p_{\lambda}H^I_{\alpha}\right)\;,\nonumber
\end{eqnarray}
where the right hand side are two-dimensional conformal diffeomorphisms. A $Q$-invariant action may be constructed by considering 
$$
\epsilon S[h,\chi,C,X,\alpha,\rho,H]=\delta_{\epsilon}V[h,X,\rho,H]\;,
$$
where $V[h,X,\rho,H]$ is diffeomorphism invariant. In this case, the action of $Q$ on $S[h,\chi,...]$ is equivalent to the action of $Q^2$ on  $V[h,X,\rho,H]$ and so, equivalent to the action of a conformal diffeomorphism on $V[h,X,\rho,H]$. Thus, if $V[h,X,\rho,H]$ is invariant under infinitesimal conformal diffeomorphisms, the action $S[h,\chi,C,X,\alpha,\rho,H]$ will be invariant under $Q$ transformations, as required. A useful choice is \cite{Witten:1988xj}
$$
V[h,X,\rho,H]=\int_{\Sigma}\rd^2 \sigma \sqrt{-h}h^{\alpha\beta}g_{IJ}(X)\rho^I_{\alpha}\left(\partial_{\beta}X^J-\frac{1}{4}H^J_{\beta}\right)\;.
$$
This choice for $V[h,X,\rho,H]$ leads to the action (where it will now be assumed that the target space is K\"{a}hler) $S=S_H+...$, where the ellipses denote $H^I_{\alpha}$-independent terms and
$$
S_H=\frac{1}{2}\int_{\Sigma}\rd^2 \sigma \sqrt{-h}g_{IJ}h^{\alpha\beta}\left(H^I_{\alpha}\partial_{\beta}X^J-\frac{1}{4}H^I_{\alpha}H^J_{\beta}\right)\;.
$$
$H^I_{\alpha}$ appears algebraically and so plays the role of a constraint field that ensures that the fermionic symmetry can be realised off-shell. Solving this constraint 
$$
H^I_{\alpha}=\p_{\alpha}X^I+J^I{}_J\varepsilon_{\alpha}{}^{\beta}\p_{\beta}X^J\;,
$$
where the self-duality condition $\rho^I_{\alpha}=J^I{}_J\varepsilon_{\alpha}{}^{\beta}\rho^J_{\beta}$ has been used and inserting this expression for $H^I_{\alpha}$ back into the action, gives an action  for a topological sigma model coupled to topological gravity \cite{Witten:1988xj}
\begin{eqnarray}\label{sigma}
S&=&\int_{\Sigma}\rd^2 \sigma \sqrt{-h}\left(\frac{1}{2}g_{IJ}h^{\alpha\beta}\partial_{\alpha}X^I\partial_{\beta}X^J+\frac{1}{2}\varepsilon^{\alpha\beta}J_{IJ}\partial_{\alpha}X^I\partial_{\beta}X^J-ig_{IJ}\rho^I_{\alpha}\left(h^{\alpha\beta}\nabla_{\beta}\alpha^J-\chi^{\alpha\beta}\p_{\beta}X^J\right)\right.\nonumber\\
&&\left.+\frac{1}{8}h^{\alpha\beta}R_{IJKL}\alpha^I\alpha^J\rho^K_{\alpha}\rho^L_{\beta}-\frac{i}{8}\varepsilon^{\alpha\beta}g_{IJ}\rho^I_{\alpha}\rho^J_{\beta}\varepsilon^{\rho\lambda}\left(\nabla_{\rho}C_{\lambda}+i\chi_{\rho\sigma}\chi^{\sigma}{}_{\lambda}\right)-\frac{i}{4}g_{IJ}h^{\alpha\beta}\rho^I_{\alpha}C^{\lambda}\nabla_{\lambda}\rho^J_{\beta}\right)\nonumber\;,
\end{eqnarray}
where
$$
\nabla_{\alpha}\alpha^I=\p_{\alpha}\alpha^I+\Gamma^I_{JK}\alpha^J\p_{\alpha}X^K\;.
$$
Truncating out the $C_{\alpha}$ fields to leaves a theory in which the $Q$ symmetry acts on the gravitational sector as
$$
\delta_{\epsilon}h_{\alpha\beta}=2i\epsilon\chi_{\alpha\beta}\;,	\qquad	\delta_{\epsilon}\chi_{\alpha\beta}=0
$$
and now $Q^2=0$. The algebra is a contraction of the algebra (\ref{Qsquared}) generated by the conformal and $Q$-transformations\footnote{A useful result in showing this is
$$
\delta_{\epsilon}\Gamma^{\lambda}_{\alpha\beta}=i\epsilon h^{\lambda\sigma}\left(\nabla_{\alpha}\chi_{\beta\sigma}+\nabla_{\beta}\chi_{\alpha\sigma}-\nabla_{\sigma}\chi_{\alpha\beta}\right)\;,
$$
which satisfies $\delta_{\epsilon}\delta_{\tilde{\epsilon}}\Gamma^{\lambda}_{\alpha\beta}=0$, where $\Gamma^{\lambda}_{\alpha\beta}$ is a \emph{world-sheet} Levi-Civita connection.}. Making use of the notation $\delta_{W}(\zeta)$, for an infinitesimal transformation with generator $W$ and parameter $\zeta$, the algebra of the truncated theory may be written as 
$$
[\delta_C(\xi),\delta_C(\tilde{\xi})]=\delta_C(\nu)\;,	\qquad	[\delta_C(\xi),\delta_Q(\varepsilon)]=0\;,	\qquad	[\delta_Q(\epsilon),\delta_Q(\varepsilon)]=0\;.
$$
where $C$ is a generator of conformal diffeomorphisms and $\nu^{\alpha}=2\xi^{[\alpha}\partial_{\beta}\tilde{\xi}^{\beta]}$. The contraction is given by rescaling $(\delta_C,\delta_Q)\rightarrow(\delta_C,\Lambda^{-1}\delta_Q)$ and then taking $\Lambda\rightarrow 0$. This has the same effect as truncating out the $C_{\alpha}$ fields in the action (\ref{sigma}). The resulting sigma model with this, now nilpotent, $Q$ symmetry has action
\begin{eqnarray}\label{witten}
S&=&\int_{\Sigma}\rd^2 \sigma \sqrt{-h}\left(\frac{1}{2}g_{IJ}h^{\alpha\beta}\partial_{\alpha}X^I\partial_{\beta}X^J+\frac{1}{2}\varepsilon^{\alpha\beta}J_{IJ}\partial_{\alpha}X^I\partial_{\beta}X^J-ig_{IJ}\rho^I_{\alpha}\left(h^{\alpha\beta}D_{\beta}\alpha^J-\chi^{\alpha\beta}\p_{\beta}X^J\right)\right.\nonumber\\
&&\left.+\frac{1}{8}h^{\alpha\beta}R_{IJKL}\alpha^I\alpha^J\rho^K_{\alpha}\rho^L_{\beta}\right)\;.
\end{eqnarray}
It is interesting to note that this model, with the $C_{\alpha}$'s truncated out, bears a close resemblance to the conventional (1,1) ten-dimensional superstring theory, with $\chi^{\alpha\beta}$ playing the role of the world-sheet gravitino.

\subsection{Half-Twisted Sigma Models}

From now on it will be assumed that the target space is K\"{a}hler. The theory (\ref{witten}) has a (2,2) symmetry which may be made manifest by using complex coordinates for the target space and light-cone coordinates on the world-sheet. The metric and K\"{a}hler form are then given by $\rd s^2=g_{\bar{\iota}j}\rd X^{\bar{\iota}}\rd X^j$ and $J=\frac{1}{2}g_{\bar{\iota}j}\rd X^{\bar{\iota}}\wedge \rd X^j$, where $X^I=(X^{\bar{\iota}},X^j)$ etc. Momentarily ignoring topological gravity by setting $h_{\alpha\beta}=\eta_{\alpha\beta}$ and $\chi_{\alpha\beta}=0$, the action (\ref{witten}) becomes
\begin{equation}\label{22}
S=\int_{\Sigma}\rd^2 \sigma\; \left(g_{\bar{\iota}j}\partial_+ X^j\partial_-X^{\bar{\iota}}-ig_{\bar{\iota}j}\rho_+^{\bar{\iota}}\nabla_-\alpha^j-ig_{\bar{\iota}j}\rho_-^j\nabla_+\alpha^{\bar{\iota}}+\frac{1}{8}R_{i\bar{\iota}k\bar{k}}\alpha^i\alpha^{\bar{\iota}}\rho_-^k\rho_+^{\bar{k}}\right)\;,
\end{equation}
where $\rd^2 \sigma=\rd\sigma^+\rd\sigma^-$. This action has the (2,2) symmetry, explicit transformations for which may be found in \cite{Witten:1991zz}. Of particular interest is a (2,0) subgroup generated by $Q$ and $\widetilde{Q}_-$. The $Q$-transformations are given infinitesimally by
\begin{equation}\label{Q}
\delta_{\epsilon}X^j=0\;,	\qquad	\delta_{\epsilon}X^{\bar{\iota}}=i\epsilon\alpha^{\bar{\iota}}\;,	\qquad	\delta_{\epsilon}\alpha^{\bar{\iota}}=0\;,	\qquad	\delta_{\epsilon}\rho_-^j=\epsilon\partial_-X^j\;,
\end{equation}
and the symmetry generated by $\widetilde{Q}_-$ with vector parameter $\tilde{\epsilon}=\tilde{\epsilon}^-\p_-$ is given by
\begin{equation}\label{Qtilde}
\delta_{\tilde\epsilon}X^j=i\tilde{\epsilon}^-\rho_-^j\;,	\qquad	\delta_{\tilde\epsilon}X^{\bar{\iota}}=0\;,	\qquad	\delta_{\tilde\epsilon}\alpha^{\bar{\iota}}=\tilde{\epsilon}^-\partial_-X^{\bar{\iota}}\;,	\qquad	\delta_{\tilde\epsilon}\rho_-^j=0\;.
\end{equation}
These infinitesimal transformations satisfy
$$
Q^2=0\;,	\qquad	\widetilde{Q}_-^2=0\;,	\qquad	\{Q,\widetilde{Q}_-\}=i\partial_-\;.
$$
There is also an obvious (0,2) symmetry, which when combined with the (2,0) symmetry generated by $Q$ and $\widetilde{Q}_-$ generates the full (2,2) symmetry. A sigma model with only the (2,0) symmetry is given by a truncation of (\ref{22}) by setting the $\alpha^i$ and $\rho^{\bar{\iota}}_-$ fields to zero which, by self-duality, implies that $\rho^j_+$ also vanishes \cite{Witten:1988xj}. Of particular interest will be the coupling this (2,0) sigma model to world-sheet topological gravity to give a (2,0) string theory that can be used as a basis for a Heterotic twistor string theory \cite{Mason:2007zv}.

\section{Closed Twistor String Theory}

The (2,0) sigma model was introduced in \cite{Witten:1988xj} and further investigated in \cite{Witten:2005px}. In the context of twistor string theory it was studied in \cite{Mason:2007zv} and the work presented in this article will draw heavily on the results of that paper. In this section the coupling of the sigma model presented in \cite{Mason:2007zv} to a half-twisted version of world-sheet topological gravity will be investigated.

\subsection{(2,0) Model with Topological Gravity}

The Lorentz group in two dimensions $SO(1,1)$ is abelian and, in contrast to higher dimensions, there is a meaningful Lorentz-invariant notion of left and right moving modes in the free theory. Even though the string theory of interest will not be initially defined in terms of a flat target space metric and hence the world-sheet theory will not be free, the physical observables do not depend on the choice of target space metric, and so we are free to choose the flat metric at least locally \cite{Witten:2005px}. This suggests that, at least in a patch $U\subset M$ of the target, left and right movers can be decoupled in a Lorentz-invariant way. The question then arises as to whether a fermionic symmetry can be introduced whose action on the components of the world-sheet metric depends on whether that component $h_{\alpha\beta}$ couples to left or right moving modes. One might consider such a theory to be a chiral or `half-twisted' version of topological gravity and the natural sigma model one would couple it to would be the topological cousin of the Heterotic string: the (2,0) model. Consider the action
\begin{equation}\label{action1}
S=\int_{\Sigma}\rd^2 \sigma\; e\; g_{\bar{\iota}j}\left(\partial_+ X^j\partial_-X^{\bar{\iota}}-i\rho_-^j\left(\nabla_+\alpha^{\bar{\iota}}+\chi_{++}\partial_-X^{\bar{\iota}}\right)\right)\;,
\end{equation}
where $\nabla_+\alpha^{\bar{\iota}}=\partial_+\alpha^{\bar{\iota}}+\Gamma^{\bar{\iota}}_{\bar{k}\bar{l}}\alpha^{\bar{l}}\partial_+X^{\bar{k}}$ and $e=\sqrt{-h}=\sqrt{-\det(h_{\alpha\beta})}$. This action is invariant under
$$
\delta_{\epsilon}e_+{}^{\alpha}=i\epsilon\chi_{++}e_-{}^{\alpha}\;,	\qquad	\delta_{\epsilon}e_-{}^{\alpha}=0\;,	\qquad	\delta_{\epsilon}\chi_{++}=0\;,
$$
$$
\delta_{\epsilon}X^j=0\;,	\qquad	\delta_{\epsilon}X^{\bar{\iota}}=i\epsilon\alpha^{\bar{\iota}}\;,	\qquad	\delta_{\epsilon}\alpha^{\bar{\iota}}=0\;,	\qquad	\delta_{\epsilon}\rho_-^j=\epsilon\partial_-X^j\;.
$$
Care must be taken to note that the $\pm$ indices are frame indices which, unlike coordinate indices, do transform under $Q$  so for example
$$
\delta_{\epsilon}(\p_+X^j)=\delta_{\epsilon}(e_+{}^{\alpha}\p_{\alpha}X^j)=i\epsilon\chi_{++}\p_-X^j\;,
$$
which is not generally zero. In terms of the world-sheet metric $h_{\alpha\beta}$, the fermionic transformations are
$$
\delta_{\epsilon}h_{++}=2i\epsilon\chi_{++}h_{+-}\;,	\qquad	\delta_{\epsilon}h_{+-}=i\epsilon\chi_{++}h_{--}\;,	\qquad	\delta_{\epsilon}h_{--}=0\;.
$$
These transformations are precisely what was hoped for in such a  chiral theory of topological gravity. It is clear that the world-sheet gravity is not purely topological, yet is not of the conventional type. An important consequence of this half-twisting of the world-sheet gravity is that the central charge count of the theory is altered. In particular the contributions to the conformal anomaly due to the gauge-fixing of the $h_{++}$ and $\chi_{++}$ degrees of freedom cancel. In contrast, $h_{--}$ contributes the usual $+26$ to the central charge and must be balanced by introducing other matter sectors to the theory.

Direct application of the $Q$ operator to the action (\ref{action1}) demonstrates that this theory is $Q$-invariant. Alternatively, one can see this from the fact that $Q^2=0$ and the action is $Q$-exact $S=\{Q,V\}$ where
$$
V=\int_{\Sigma}\rd^2\sigma\,e\,g_{\bar{\iota}j}\rho_-^j\partial_+X^{\bar{\iota}}
$$
The Noether current for the $Q$-symmetry is given by $J_Q=g_{\bar{\iota}j}\alpha^{\bar{\iota}}\partial_-X^j\rd\sigma^-$ which gives an expression for $Q$ in terms of the fields in the theory
$$
Q=\oint \rd\sigma^-e\, g_{\bar{\iota}j}\alpha^{\bar{\iota}}\partial_-X^j\,.
$$
A superspace description of this string theory is outlined in section 4.1. For a twistor string the target space is required to be $M=\C\P^{3}$ and in many cases it is more useful to write such a sigma model in terms of homogenous coordinates $Z^I$ (coordinates on $\C^3$) and introduce a $GL(1;\C)$ gauging $(Z^I,Z^{\bar{I}})\rightarrow (tZ^I,t^*Z^{\bar{I}})$, where $t\in\C$ cannot be zero. The action, written in terms of these homogenous coordinates becomes
$$
S= \int_{\Sigma}\rd^2\sigma \,G_{\bar{I}J}\,D_+Z^{\bar{I}}D_-Z^J+...
$$
where $D_-Z^I=\partial_-Z^I+A_-Z^I$ and $D_+Z^{\bar{I}}=\partial_+Z^{\bar{I}}+A_+Z^{\bar{I}}$ are $GL(1;\C)$ covariant derivatives and $G_{\bar{I}J}=|Z|^{-2}\delta_{\bar{I}J}$. To see that these formulations are equivalent, integrate out the gauge fields to get 
$A_+=-|Z|^{-2}Z^I\partial_+Z_I$ and $A_-=-|Z|^{-2}Z_I\partial_-Z^I$. Substituting these back in gives
$$
S=\int_{\Sigma}\rd^2\sigma\, g_{\bar{I}J}\,\partial_+Z^{\bar{I}}\partial_-Z^J+...
$$
where $g_{\bar{I}J}$ is the Fubini-Study metric on $\C\P^3$ in homogenous coordinates. The description in terms of the Fubini-Study metric in terms of coordinates $(X^j,X^{\bar{\iota}})$ on $\C\P^3$ is then straightforward. Gauge fixing the $GL(1;\C)$ symmetry requires the addition of ghost fields $(u,v)$. These ghosts contribute -2 to the central charge which balances the extra +2 coming from the target space being $\C^4$ as opposed to $\C\P^3$, so the descriptions are equivalent.

\subsubsection{Symmetries}

The main focus has thus far been on the $Q$-symmetry above. It is less obvious, but the theory is also invariant under the $\widetilde{Q}_-$ transformations (\ref{Qtilde}), provided that $\p_+\tilde{\epsilon}^-=0$. The varation of the Lagrangian under $\widetilde{Q}_-$ gives a total derivative and thus the action (\ref{action1}) is invariant. In addition to the $Q$ and $\widetilde{Q}_-$ symmetries, the theory is invariant under world-sheet diffeomorphisms and Weyl transformations. The stress-energy tensor $T_{\alpha\beta}$ and its $Q$-partner $G_{\alpha\beta}$, which respectively generate the conformal and $\widetilde{Q}_-$ transformations, are given by\footnote{$T_{\alpha\beta}$ and $G_{\alpha\beta}$ are defined by
$$
T_{\alpha\beta}=\frac{1}{e}\frac{\delta S}{\delta h^{\alpha\beta}}\;,	\qquad	G_{\alpha\beta}=\frac{i}{e}\frac{\delta S}{\delta \chi^{\alpha\beta}}\;.
$$}
$$
T_{++}=g_{\bar{\iota}j}\p_+X^{\bar{\iota}}\p_+X^j\;,	\qquad	T_{--}=g_{\bar{\iota}j}\left(\p_-X^{\bar{\iota}}\p_-X^j-i\rho_-^j\p_-\alpha^{\bar{\iota}}\right)\;,
$$
and
$$
G_{++}=0\;,	\qquad	G_{--}=g_{\bar{\iota}j}\rho_-^j\p_-X^{\bar{\iota}}\;.
$$
It is not hard to show that $\widetilde{Q}_-$ may be written as a contour integral
$$
\widetilde{Q}_-=\oint\rd\sigma^-G_{--}\;.
$$
Note that $T_{--}$ is $Q$-exact
$$
T_{--}=\{Q,G_{--}\}\;,
$$
and $T_{++}$ is $Q$-closed on-shell. The fact that only $T_{--}$ is trivial in $Q$-cohomology and not the full stress tensor highlights the fact that this is not a fully topological theory \cite{Witten:2005px}, as the physics does depend on \emph{some} component of the world-sheet metric, specifically on the component $h_{++}$. In order to have meaningful path integral expressions for correlation functions, it will be necessary to gauge fix the conformal and $\widetilde{Q}_-$ symmetries. This issue will be considered using the standard BRST procedure in section 3.2.

The theory has a $U(1)$ $R$-symmetry with current
$$
J_R=g_{\bar{\iota}j}\alpha^{\bar{\iota}}\rho_-^j\rd\sigma^-\;,
$$
in which the fields $X^{\bar{\iota}}$, $X^j$, $\alpha^{\bar{\iota}}$, and $\rho_-^j$ have $R$-charges $0$, $0$, $+1$, and $-1$ respectively. It is clear from the  transformations (\ref{Q}) that the generator of the fermionic symmetry $Q$ has $R$ charge $+1$. The conservation of the charge associated with this current (broadly speaking the difference between the number of $\alpha^{\bar{\iota}}$ and $\rho^j_-$ zero modes) imposes certain constraints on non-trivial correlation functions and will be discussed further in section 3.3.

\subsubsection{Additional Fields}

So far the discussion has focussed on the (2,0) theory with a general K\"{a}hler background. Of particular interest is the coupling of the Heterotic twistor sigma model introduced in \cite{Mason:2007zv} to topological gravity. There is no obstruction to choosing the target space to be $\C\P^3$ which describes the bosonic part of twistor space but one would really like to describe the full $\C\P^{3|4}$ of supertwistor space. In \cite{Mason:2007zv}, it was shown how to include the target space fermionic coordinates by introducing fermionic world-sheet scalars $\psi_a$ and fermionic  one-forms $\bar{\psi}^a_+$ with action
$$
S_{\bar{\psi},\psi}=-i\int_{\Sigma}\rd^2 \sigma\; e\;\bar{\psi}_+^a\p_-\psi_a\;.
$$
These $\psi$'s take values in the pull-back of the bundle ${\cal O}(1)^{\oplus 4}$ and provide a way to incorporate the fermionic directions of the supertwistor target space $\C\P^{3|4}$.

Gauge degrees of freedom are incorporated by  introducing world-sheet fermions $\lambda_m$ ($m=1,2,...N$), coupling to the pull-back to $\Sigma$ of a background (0,1)-form ${\cal A}(X)={\cal A}_{\bar{\iota}}\rd X^{\bar{\iota}}$ with action
$$
S_{\bar{\lambda},\lambda}=-i\int_{\Sigma}\rd^2 \sigma\; e\,\bar{\lambda}^m\Gamma_+{\cal D}_-\lambda_m+\int_{\Sigma}\rd^2 \sigma\; e\;F_{\bar{\iota}jm}{}^n\bar{\lambda}^m\Gamma_+\lambda_n\rho^j_-\alpha^{\bar{\iota}}
$$
where ${\cal D}_-\lambda^m=\partial_-\lambda^m+\lambda^n{\cal A}_{\bar{\iota}n}{}^m(X)\partial_-X^{\bar{\iota}}$ and $F_m{}^n=F_{\bar{\iota}jm}{}^n\rd X^{\bar{\iota}}\wedge\rd X^j$ is the (1,1) curvature of the (0,1)-form connection ${\cal A}_{\bar{\iota}m}{}^n\rd X^{\bar{\iota}}$. $\Gamma_+$ is a two-dimensional gamma matrix and world-sheet spinor indices have been suppressed. As with the conventional heterotic string, there is a rigid world-sheet symmetry
$$
\lambda^m\rightarrow U^m{}_n\lambda^n\;,	\qquad	\bar{\lambda}_m\rightarrow \bar{\lambda}_n(U^{-1})_m{}^n\;,	\qquad	{\cal A}_{\bar{\iota}m}{}^n\rightarrow U_m{}^p{\cal A}_{\bar{\iota}p}{}^q(U^{-1})^q{}_n-\p_{\bar{\iota}}U_m{}^p(U^{-1})_p{}^n
$$
where $U^m{}_n$ takes values in the Lie algebra of a group $G$, which for our purposes we shall assume is $SU(2N)$ where $N$ is an integer to be fixed by conformal invariance.  We allow $U^m{}_n$ to depend  $X^{\bar{\iota}}$ but not on the world-sheet coordinates so that the symmetry is a rigid transformation on the world-sheet but a gauge symmetry in the target space.

Putting this all together, the action for the Heterotic twistor string is given by
\begin{eqnarray}
S&=&\int_{\Sigma}\rd^2 \sigma\;e \left(g_{\bar{\iota}j}\partial_+ X^j\partial_-X^{\bar{\iota}}-i\bar{\psi}^a_+\p_-\psi_a-ig_{\bar{\iota}j}\rho_-^j\left(\nabla_+\alpha^{\bar{\iota}}+\chi_{++}\partial_-X^{\bar{\iota}}\right)\right.\nonumber\\
&&\left.-i\bar{\lambda}^m\Gamma_+{\cal D}_-\lambda_m+F_{m\bar{\iota}j}{}^n\bar{\lambda}_n\Gamma_+\lambda^m\rho^j_-\alpha^{\bar{\iota}}\right)
\end{eqnarray}
In accordance with the Penrose-Ward transform \cite{Ward:1977ta}, it will be assumed that the gauge field is flat when pulled back to the world-sheet so that the last term may be ignored.

\subsection{BRST Quantization}

In order to have a sensible path integral description of correlation functions it is necessary to gauge fix the diffeomorphism, Weyl and $\widetilde{Q}_-$ symmetries. The world-sheet diffeomorphism invariance is used to gauge fix the zwiebein so that $e_+{}^+=0=e_-{}^-$. The $SO(1,1)$ Lorentz symmetry and the Weyl symmetry are then used to remove the remaining two zwiebien degrees of freedom so that the world-sheet metric can be put into the form
$$
\eta_{\alpha\beta}=\left(\begin{array}{cc}
0 & 1 \\
1 & 0
\end{array}\right)
$$
This gauge-fixing requires the introduction of reparameterisation ghosts $c=c^{\alpha}\partial_{\alpha}$ and traceless, symmetric, anti-ghosts $b_{\alpha\beta}$ with action
$$
S_{b,c}=\int_{\Sigma}\rd^2\sigma\left(b_{++}\p_- c^++\tilde{b}_{--}\p_+\tilde{c}^-\right)\;,
$$
and the $\widetilde{Q}_-$ symmetry is fixed by the introduction of the right-moving beta-gamma system
$$
S_{\beta,\gamma}=\int_{\Sigma}\rd^2\sigma\tilde{\beta}_{--}\p_+\tilde{\gamma}^-\;.
$$
$b_{\alpha\beta}$ and $c^{\alpha}$ are fermonic, whilst $\tilde{\beta}_{--}$ and $\tilde{\gamma}^-$ are bosonic. The resulting gauge-fixed action is
\begin{eqnarray}\label{action}
S&=&\int_{\Sigma}\rd^2 \sigma\; \left(g_{\bar{\iota}j}\partial_+ X^{\bar{\iota}}\partial_-X^j-i\bar{\psi}^a_+\p_-\psi_a-ig_{\bar{\iota}j}\rho_-^j\nabla_+\alpha^{\bar{\iota}}-i\bar{\lambda}^m\Gamma_+{\cal D}_-\lambda_m\right.\nonumber\\
&&\left.+b_{++}\,\p_-c^++\tilde{b}_{--}\,\p_+\tilde{c}^-+\tilde{\beta}_{--}\,\p_+\tilde{\gamma}^-\right)\;,
\end{eqnarray}
which is invariant under the BRST transformations generated by the BRST charge $Q_{\text{BRST}}$, which includes the stress tensor and supercurrent. One could also consider a Liouville sector which would play the role of a gauge-fixing term for the Weyl transformations. Such terms, and the important role they play in two-dimensional topological gravity, were considered in \cite{Verlinde:1990ku,Dijkgraaf:1990qw}. Liouville terms will not be introduced here, although it would be interesting to explicitly incorporate them into a description of the twistor string. Consider the nilpotent operator ${\cal Q}=Q+Q_{\text{BRST}}$. Using standard arguments\footnote{The symmetry under consideration is replaced by a symmetry in which ghosts play the role of the parameters. Specifically we have
$$
\zeta^{\alpha}\rightarrow c^{\alpha}\;,	\qquad	\tilde{\epsilon}^-\rightarrow\tilde{\gamma}^-\;.
$$} for $Q_{\text{BRST}}$, the operator ${\cal Q}={\cal Q}_L+{\cal Q}_R$ is given by
$$
{\cal Q}_L=Q_{\epsilon}+\oint\rd\sigma^- \tilde{c}^-\left(T_{--}+\frac{1}{2}T_{--}^{gh}\right)+\oint\rd\sigma^- \tilde{\gamma}^-\left(G_{--}+\frac{1}{2}G_{--}^{gh}\right)\;,
$$
and
$$
{\cal Q}_R=\oint\rd\sigma^+ c^+\,\left(T_{++}+\frac{1}{2}T_{++}^{gh}\right)\;,
$$
where $Q_{\epsilon}$ is the fermionic generator $Q$ with parameter $\epsilon$ i.e. $Q_{\epsilon}=\oint\rd\sigma^-\epsilon g_{\bar{\iota}j}\alpha^{\bar{\iota}}\p_-X^j$. The structure of this formula is standard (the BRST symmetry is generated by the generator of the conformal transformation $T_{\pm\pm}$ with the $c^{\pm}$ ghost as a parameter and so on). The Cohomology of ${\cal Q}$ is equivalent to the physical spectrum. It is interesting to note that the fermionic symmetries, generated by $Q$ and $\widetilde{Q}_-$, when taken together, have a similar structure to that of the Heterotic string\footnote{The BRST charge of the Heterotic string is of the form
$$
Q_{\text{BRST}}=\oint\rd\sigma^-\tilde{c}^-\left(T_{--}+\frac{1}{2}T^{gh}_{--}\right)+\oint\rd\sigma^-\tilde{\gamma}\left(G_-+\frac{1}{2}G_-^{gh}\right)
$$
where $\tilde{\gamma}$ are now ghosts for the gravitino modes and the supercurrent is
$$
G_{-}=g_{\mu\nu}\psi^{\mu}\p_-X^{\nu}=g_{\bar{\iota}j}\left(\psi^{\bar{\iota}}\p_-X^j+\psi^j\p_-X^{\bar{\iota}}\right)\;,
$$
where the world-sheet spinor indices have been suppressed.}
$$
Q_{\epsilon}+\oint\rd\sigma^-\,\tilde{\gamma}^-\,G_{--}=\oint\rd\sigma^- g_{\bar{\iota}j}\left(\epsilon\alpha^{\bar{\iota}}\p_-X^j+\tilde{\gamma}^-\rho^j_-\p_-X^{\bar{\iota}}\right)
$$
so that much of the formal structure of BRST quantization of the Heterotic string may be applied to the half-twisted topological string\footnote{The $G_{--}$ condition requires the observables to be $\bar{\p}^{\dagger}$-closed. It would be interesting to understand this condition from the perspective of twistor theory a little better.}. This is not surprising given the close relationship between these sigma models. Care must be taken to ensure that the total central charge for the gauge-fixed theory vanishes. Using the standard formula for the contribution $c_{\ell}$ to the conformal anomaly for free fields with a first order action (see for example \cite{Lust})
$$
c_{\ell}=-\varepsilon(12\ell^2-12\ell+2)\;,
$$
where $\varepsilon=+1$ for fields with Fermi statistics and $\varepsilon=-1$ for fields with Bose statistics. The fields of the first order system have conformal weight $\ell$ and $1-\ell$, the cancellation of central charge contributions may be checked for left and right moving sectors\footnote{For the ($b_{\alpha\beta},c^{\alpha}$) system and also for the ($\tilde{\beta}_{--},\tilde{\gamma}^-$) system, $\ell=2$. $\ell=1$ for the ($\alpha^{\bar{\iota}},\rho_-^j$) and ($\bar{\psi}_+,^a\psi^a$) systems and $\ell=1/2$ or the ($\bar{\lambda}^m,\lambda^m$) system.}. These are all free fields so the net central charge is given by the sum of the central charges for each component\footnote{The observables do not depend on the local form of the metric as this enters into the action as a $Q$-exact term so, in patch $U\subset\C\P^3$, we can choose a flat metric $g_{\bar{\iota}j}=\delta_{\bar{\iota}j}$. The scalar theory is free in this case so the central charge count above is correct. This also gives some justification of the abuse of notation in calling the modes of the theory with non-flat target space left- and right-movers.}. The result is summarised in table 1 below. The key points are that the anomaly vanishes for $N=28$ as found in \cite{Witten:2003nn,Berkovits:2004hg,Mason:2007zv}, and also that topological gravity in the right-moving sector ensures that no additional fields need to be introduced for the central charge to vanish.

\begin{table}[htdp]
\caption{Twistor String Field Content}
\begin{center}
\begin{tabular}{|c c c c | c c c c|}
\hline
&  Left-Movers & & &  & Right-Movers & & \\
\hline\hline
Field & Central Charge & Spin & Statistics & Field & Central Charge & Spin & Statistics \\
\hline
&&&&&&&\\
$\p_-X^j$		&	$(D,0)$ & 0 & \text{Bose} & $\p_+X^{\bar{\iota}}$		&	$(0,D)$ & 0 & \text{Bose}	\\
$\psi^a_+,\psi_a$	&	$(-D-2,0)$	& 1, 0 & \text{Fermi}  & $\alpha^{\bar{\iota}}$, $\rho^j_-$	&	$(0,-D)$	& 0, 1 & \text{Fermi} \\
$b_{++}$, $c^+$	&	$(-26,0)$ & 2,1 & \text{Fermi} & $\tilde{b}_{--}$, $\tilde{c}^-$	&	$(0,-26)$ & 2,1 & \text{Fermi}	\\
$\lambda^m$	&	$(N,0)$ & $\frac{1}{2}$ & \text{Fermi} & $\tilde{\beta}_{--}$, $\tilde{\gamma}^-$	&	$(0,+26)$ & 2, 1 & Bose\\
&&&&&&&\\
\hline
\end{tabular}
\end{center}
\label{default}
\end{table}%

\subsection{Anomalies}

The presence of first order fermion terms in the action with U(1) symmetries leads to a potential ambiguity in the path integral definition of correlation functions. As described in \cite{Moore:1984ws}, a the path integral will not be a function but will be a section of a line bundle. If the (Quillen) connection of this bundle is flat, then there is no ambiguity in choosing a section and the path integral is well-defined. This was discussed at length for the Heterotic twistor string in \cite{Mason:2007zv}. A useful way to think about this anomaly is in terms of the conservation of the charge associated to the Noether current of the U(1) symmetry.

Considering first world-sheet fields that do not carry target space indices, anomalies arise due to an imbalance in the number of zero modes in the first order system with action of the form $\int i\Phi\p_{\pm}\Psi$, where $\Phi$ has conformal weight $\ell$ and $\Psi$ has weight $1-\ell$ (see for example chapter 13 of \cite{Lust}). The anomaly due to this system vanishes if
$$
N_{\Phi}-N_{\Psi}=(1-2\ell)(g-1)
$$
is zero, where $N_{\varphi}$ denotes the number of zero modes of the field $\varphi$. For example, the $SU(N)$ gauge spinors have a current $\bar{\lambda}^m\Gamma_+\lambda_m\rd\sigma^+$. For these fields $\ell=1/2$ and so the anomaly vanishes automatically. In addition there are global U(1) ghost symmetries given by $b_{\alpha\beta}\rightarrow e^{-i\varphi}b_{\alpha\beta}$ and $c^{\alpha}\rightarrow e^{i\varphi}c^{\alpha}$, for some phase $\varphi$ with current\footnote{Normal ordering is implicit.}
$$
j=c^+b_{++}\rd\sigma^+\;,	\qquad	\tilde{j}=\tilde{c}^-\tilde{b}_{--}\rd\sigma^-\;.
$$
Similarly there is a $Q$-ghost current
$$
\tilde{k}=\tilde{\gamma}^-\tilde{\beta}_{--}\rd\sigma^-\;.
$$
In both of these cases $\ell=2$ and the anomaly for the vacuum theory, with no further insertions of these fields, is $3-3g$ for a world-sheet of genus $g$. As mentioned above, an interesting issue is the similarity between this Heterotic twistor string and the conventional ten-dimensional supersymmetric Heterotic theory. In both cases the $(b,c)$ and $(\beta,\gamma)$ ghosts have fermionic and bosonic statistics respectively; however, in contrast to the action of supersymmetry, the $Q$ symmetry does not change the spin of the fields it acts upon (it is a scalar operator). As such, the conformal ghosts and $Q$-ghosts of the twistor string have the same spin. In the conventional Heterotic string the conformal and superconformal ghosts have differing spins and the vacuum anomaly associated with the superconformal ghosts is $2-2g$. These $N_b-N_c=3-3g$ anomalies show that the ghost and $Q$-ghost vacua have a ghost number charge $3-3g$. This charge must be balanced by operator insertions on the world-sheet for a correlation function to be well-defined and this will be important when scattering amplitudes are considered.

The $\rho^j_-$, $\alpha^{\bar{\iota}}$, $\bar{\psi}^a_+$ and $\psi_a$ fields are sections of pull-backs of bundles over the target space, so the anomaly analysis requires a careful consideration of these bundles. Such an analysis may be found in \cite{Mason:2007zv} and only the results will be repeated here. The ($\rho^j_-$, $\alpha^{\bar{\iota}}$) system is anomalous unless
$$
N_{\alpha}-N_{\rho}=4d+3(1-g)\;,
$$
vanishes, where $d$ is the degree of the map $X:\Sigma\rightarrow\C\P^{3|4}$. Similarly, the ($\bar{\psi}^a_+$, $\psi_a$) system is anomalous unless
$$
N_{\psi}-N_{\bar{\psi}}=4(d+1-g)\;.
$$
vanishes. These results are modified by operator insertions in the obvious way. For example, for an $n$-point N$^k$MHV gluon amplitude: the $k+2$ negative helicity gluons (elements of $H^{0,1}(\C\P^3;{\cal O}(-4))$), which each contribute $4(k+2)$ $\psi's,$; and  $n-k-2$ positive helicity gluons (elements of $H^{0,1}(\C\P^3;{\cal O})$) which does not contribute any additional powers of $\psi$, modify the anomaly cancellation condition so that the degree of the map $X:\Sigma\rightarrow\C\P^3$ is required to be \cite{Witten:2003nn}
$$
d=k+1+g\;.
$$
Similar results hold for scattering amplitudes involving other fields.

\subsection{Holomorphy of Maps}

It is well known in twistor string theory that the path integral of the twistor string localises on holomorphic maps of $\Sigma$ into twistor space. These arguments are briefly reviewed here as they will be important later on. More details may be found in \cite{Witten:2005px}. In section 3.3.1 it was shown that the matter stress tensor is Q-closed but importantly, $T_{--}$ is also $Q$-exact; $T_{--}=\{Q,G_{--}\}$. Anti-holomorphic world-sheet translations are generated by\footnote{Recalling that $\{Q,\widetilde{Q}\}=\p_-$, we can see that $\oint\rd\sigma^-T_{--}$ generates shifts in $\sigma^-$ as follows
$$
\p_-=\{Q,\widetilde{Q}_-\}=\{Q,\oint\rd\sigma^-G_{--}\}=\oint\rd\sigma^-\{Q,G_{--}\}=\oint\rd\sigma^-T_{--}\;.
$$}
$$
\bar{L}_{-1}=\oint\rd\sigma^-T_{--}
$$
So that, for some operator ${\cal O}$, a translation in $\sigma^-$ may be written as $[\bar{L}_{-1},{\cal O}]=\partial_-{\cal O}$. The $Q$-exactness of $T_{--}$ implies that there is a $W_{-1}$ such that $\bar{L}_{-1}=\{Q,W_{-1}\}$ and thus
$$
\partial_-{\cal O}=[\bar{L}_{-1},{\cal O}]=[\{Q,W_{-1}\},{\cal O}]=\{Q,[W_{-1},{\cal O}]\}
$$
so that $\partial_-{\cal O}$ is trivial in cohomology and the physical information is holomorphic. Another argument runs as follows. Under the rescaling $(\sigma^+,\sigma^-)\rightarrow (t\,\sigma^+,\tilde{t}\,\sigma^-)$, generated by $(L_0,\bar{L}_0)$, a weight $(n,m)$ field transforms as ${\cal O}^{(n,m)}\rightarrow t^n\tilde{t}^m{\cal O}^{(n,m)}$. Again the $Q$-exactness of $T_{--}$ implies that there is a $W_0$ such that $\bar{L}_0=\{Q,W_0\}$. By definition, the action of $\bar{L}_0$ on an operator of weight $(n,m)$ is
$$
[\bar{L}_0,{\cal O}]=m{\cal O}=[\{Q,W_0\},{\cal O}]
$$
Using $[\{Q,W_0\},{\cal O}]-\{Q,[W_0,{\cal O}]\}=\{W_0,[Q,{\cal O}]\}=0$ where the last equality arises from the fact that ${\cal O}$ is an observable. We see then that if $m\neq 0$ then there exists a ${\cal W}$ such that
$$
{\cal W}=\frac{1}{m}[W_0,{\cal O}]
$$
where ${\cal O}=\{Q,{\cal W}\}$ and so, for $m\neq 0$ ${\cal O}$ is $Q$-trivial. The corresponding result is that $m=0$ for non-trivial operators ${\cal O}$.

\section{Scattering Amplitudes}

As a pre-curser to studying scattering amplitudes in twistor string theory, it is helpful to consider the partition function description of correlation functions for the conventional bosonic string in terms of an integral over moduli space. Recall the bosonic string with action
$$
S_m[X]=\int_{\Sigma}\rd^2\sigma\,\sqrt{h}\,h^{\alpha\beta}g_{\mu\nu}\p_{\alpha}X^{\mu}\p_{\beta}X^{\nu}\;.
$$
Gauge-fixing of the world-sheet metric leads to the introduction of a Fadeev-Popov conformal ghost sector
$$
S_{gh}[b,c]=\int_{\Sigma}\rd^2\sigma\,\sqrt{h}\,h^{\alpha\lambda}b_{\alpha\beta}\p_{\lambda}c^{\beta}
$$
so that the action $S[X,b,c]=S_m[X]+S_{gh}[b,c]$ is invariant under the BRST symmetry given by the usual conformal transformations with parameter $c^{\alpha}$.  In bosonic string theory one integrates over the moduli space of  $n$-pointed genus $g$ Riemann surfaces $\Sigma$ that are not related by two-dimensional diffeomorphisms $\sigma^{\alpha}\rightarrow \sigma+\xi^{\alpha}(\sigma,\bar{\sigma})$ and Weyl rescalings $h_{\alpha\beta}\rightarrow e^{\Omega}h_{\alpha\beta}$, where the diffeomorphisms are required to vanish at the $n$ punctures. Choosing a reference metric
$$
\eta_{\alpha\beta}=\left(\begin{array}{cc}
0 & 1 \\
1 & 0
\end{array}\right)\;,
$$
a general metric $h_{\alpha\beta}$, which can be written as
$$
h_{\alpha\beta}=\eta_{\alpha\beta}+\mu_{\alpha\beta} \,
$$
is related to $\eta_{\alpha\beta}$ by diffeomorphisms, Weyl scalings and changes in the moduli $m^K$, which parameterise the $6g-6+2n$ real dimensional moduli space of an $n$-pointed, genus $g$ Riemnn surface ${\cal M}_{g,n}$. The changes in the moduli are generated by vectors $\psi=\psi^K\frac{\partial}{\partial m^K}\in T{\cal M}_{g,n}$. Introducing the inner product $\langle\;,\;\rangle$ one can consider the dual space of cotangent vectors $\mu=\mu_K\rd m^K\in T^*{\cal M}_g$. These $\mu_K$ are the Beltrami differentials and give a basis for changes in the metric, modulo Weyl and diffeomorphism transformations,
$$
\mu_{\alpha\beta}=\sum_K\frac{\p h_{\alpha\beta}}{\p m^K}\rd m^K \equiv \sum_K\mu_{\alpha\beta K}\rd m^K
$$
In the standard BRST procedure the world-sheet metric $h_{\alpha\beta}$ is fixed and ghosts $c^{\alpha}$ and anti-ghosts $b_{\alpha\beta}$ are introduced via the standard Fadeev-Popov procedure. The gauge symmetry of the original theory is then replaced by the BRST symmetry of the gauge-fixed theory with the BRST-invariant action $S[X,b,c]=S_m[X]+S_{gh}[b,c]$. In the standard BRST approach the metric does not change under a BRST transformation. Following \cite{Witten:2012bh} it can be useful to allow the gauge-fixed metric to vary in some unspecified way by the action of the BRST generator. In particular, if we take the somewhat unusual step of augmenting the standard BRST transformation with
\begin{equation}\label{BRST}
\delta h_{\alpha\beta}=0	\qquad	\rightarrow	\qquad	\hat{\delta} h_{\alpha\beta}=\mu_{\alpha\beta}\;,
\end{equation}
then, under such a variation, the action changes as\footnote{This follows from the definition of the stress tensor as the variation of the action induced by a variation of the word-sheet metric.}
$$
\hat{\delta} S=\int_{\Sigma}\rd^2\sigma\sqrt{h}\,\mu_{\alpha\beta}T^{\alpha\beta}
$$
where $T^{\alpha\beta}$ is the full stress tensor for the matter \emph{and} ghost systems. Thus if instead of the usual BRST transformations, we require (\ref{BRST}) and in addition that $\hat{\delta} \mu_{\alpha\beta}=0$ and $\hat{\delta} b_{\alpha\beta}=T_{\alpha\beta}$, then an invariant action is given by \cite{Witten:2012bh}
$$
\widehat{S}=S[X,b,c]-\int_{\Sigma}\rd^2\sigma\sqrt{h}\,\mu^{\alpha\beta}\,b_{\alpha\beta}\;.
$$
This action is not only well-defined at a point on the moduli space but is invariant on the whole moduli space and so one can sensibly include it in expressions in which ${\cal M}_{g,n}$ is integrated over. The zero modes of the $b_{\alpha\beta}$ generate deformations in the metric, whilst the the zero modes of the $c^{\alpha}$ are conformal Killing vectors, so the difference $N_b-N_c=6g-6+2n$ is the number of moduli. Expanding the invariant action in terms of the moduli gives
$$
\widehat{S}=S[X,b,c]-\sum_{K=1}^{6g-6+2n}\rd m^K\int_{\Sigma}\rd^2\sigma\sqrt{h}\,\mu_{\alpha\beta K}\,b^{\alpha\beta}
$$
Consider the correlation function given by the path integral with operator insertions ${\cal O}_i$ at the point $\sigma_i^{\alpha}$ on the $n$-pointed genus $g$ Riemann surface $\Sigma$. At a point $m^K$ on ${\cal M}_{g,n}$, the functional integral
$$
{\cal F}_g({\cal O}_1...{\cal O}_n)=\int {\cal D}(X,b,c)\,{\cal O}_1(\sigma_1)...{\cal O}_n(\sigma_n)\,\exp\left(-S[X,b,c]+\sum_{K=1}^{6g-6+2n}\rd m^K\int_{\Sigma}\rd^2\sigma\sqrt{h}\,\mu_{\alpha\beta K}b^{\alpha\beta}\right)\;,
$$
can be constructed, where ${\cal F}_g({\cal O}_1...{\cal O}_n)$ is treated as a top form on ${\cal M}_{g,n}$ which is then integrated to give the correlation function
$$
\langle{\cal O}_1...{\cal O}_n\rangle_g=\int_{{\cal M}_{g,n}}F_g({\cal O}_1...{\cal O}_n)\;.
$$
Since the $b_{\alpha\beta}$ are fermionic, the exponential term in ${\cal F}_g({\cal O}_1...{\cal O}_n)$ may be expanded out to give
$$
{\cal F}_g({\cal O}_1...{\cal O}_n)=\int {\cal D}(X,b,c)\;e^{-S[X,b,c]}{\cal O}_1(\sigma_1)...{\cal O}_n(\sigma_n)\; \prod_{K=1}^{6g-6+2n}\rd m^K\int_{\Sigma}\rd^2\sigma\sqrt{h}\mu_{\alpha\beta K}b^{\alpha\beta}\;,
$$
which explicitly shows ${\cal F}_g$ to be a top form. It is useful to recall the shorthand notation for the inner product
 $$
 \oint_{\Sigma}\rd^2\sigma\sqrt{h}\mu_{\alpha\beta K}b^{\alpha\beta}:=\langle \mu_K,b\rangle\;.
 $$
The $b$ ghosts are fermionic, and so the presence of the $\langle\mu_K,b\rangle$ term ensures the integral is supported on $\delta(\langle\mu_K,b\rangle)$ so that the correlation functions may be written as
$$
\langle{\cal O}_1...{\cal O}_n\rangle_g=\int_{{\cal M}_{g,n}}\prod_{K=1}^{6g-6+2n}\rd m^K\,\int {\cal D}(X,b,c)\,\delta(\langle\mu_K,b\rangle)\;e^{-S[X,b,c]}\,{\cal O}_1(\sigma_1)...{\cal O}_n(\sigma_n)\;.
$$
At tree level ($g=0$), the above correlation function is only meaningful for $n\geq 3$ and these three punctures fix the conformal symmetry on the $\Sigma=\C\P^1$. There are no moduli in this case and so no $b_{\alpha\beta}$'s.

\subsection{Superspace Construction}

In this section a superspace inspired construction of the Heterotic twistor string is introduced. In place of the usual supercoordinates of the Heterotic superstring one may introduce the fermonic coordinates $(\theta,\theta^-)$, where $\theta$ is an anti-commuting scalar and $\theta^-$ is an anti-commuting vector. Using these coordinates, superfields which contain the twistor string world-sheet fields may be constructed
$$
\Phi^j=X^j+i\theta^-\rho^j_-\;,	\qquad	\Phi^{\bar{\iota}}=X^{\bar{\iota}}+i\theta\alpha^{\bar{\iota}}\;,
$$
and the superspace derivative
$$
D_-=\frac{\p}{\p\theta^-}+i\theta \p_-\;,
$$
in introduced, so that 
$$
D_-\Phi^j=i\rho_-^j+i\theta\p_-X^j\;,	\qquad	\p_+\Phi^{\bar{\iota}}=\p_+X^{\bar{\iota}}+i\theta\p_+\alpha^{\bar{\iota}}\;.
$$
Of crucial importance is the fact that the vector $\theta^-$ drops out of these expressions so a $\theta^-$-independent action can be constructed
$$
S=\int_{\Sigma}\rd^2\sigma{\cal L}\;,	\qquad	{\cal L}=-i\int\rd\theta E g_{\bar{\iota}j}(\Phi)D_-\Phi^j\p_+\Phi^{\bar{\iota}}\;,
$$
where, in the analogue of Wess-Zumino gauge for the Heterotic superstring \cite{Evans:1986wt,Evans:1986ada}, the superveilbein is
$$
E_A{}^M=\left(
\begin{array}{cc}
e_a{}^{\alpha}-\frac{i}{2}\theta\chi_{a+}e_-{}^{\alpha} & -\frac{1}{2}(\chi_{a+}+\theta^- \phi_a) \\
i\theta e_-{}^{\alpha} & 1
\end{array}
\right)\;,
$$
where $\phi_a$ is the spin connection on the world-sheet and $\chi_{a+}=(\chi_{++},0)$. The super-determinant of the supervielbein $E_A{}^M$ is
$$
E=e(1-i\theta \chi_{++})\;.
$$
In this gauge, the Lagrangian is given by
$$
e^{-1}{\cal L}=g_{{\bar{\iota}}j}(X)\p_+X^{\bar{\iota}}\p_-X^j-ig_{{\bar{\iota}}j}\rho_-^j\left(\nabla_+\alpha^{\bar{\iota}}-\chi_{++}\p_-X^{\bar{\iota}}\right)\;,
$$
which is the $Q$-invariant action (\ref{action1}). This theory can be coupled to the pull-back of a (0,1)-form in twistor space by introducing a Lie algebra valued world-sheet fermion $\lambda^m$ and pulling back a (0,1)-form ${\cal A}_{\bar{\iota}}(\Phi)\rd X^{\bar{\iota}}$ from the target space to the world-sheet. A superspace Lagrangian for these fields is
$$
{\cal L}_{\cal A}=\int\rd\theta\, E\, \bar{\cal V}_m\left(D_-{\cal V}^m+{\cal V}^n{\cal A}(\Phi)_n{}^m\right)\;,
$$
where the gauge superfield is ${\cal V}^m=\lambda^m+\theta r^m$ and $r^m$ is a constraint field which appears algebraically in the action. Since $\theta$ is a scalar, $r^m$ and $\lambda^m$ have the same spin but opposite statistics. On-shell, the constraint field $r^m$ is given by
$$
r^m=\frac{i}{2}\chi_{++}\lambda^m-ik_{qn}k^{m[n}\lambda^{p]}\alpha^{\bar{\iota}}{\cal A}_{\bar{\iota} p}{}^q\;,
$$
where $k_{mn}$ is an invariant metric for the Lie group $G$. Integrating out the $\theta$, gives the contribution to the action from this $\lambda^m$ sector as
$$
e^{-1}{\cal L}=-i\bar{\lambda}_m\left(\p_-\lambda^m+\lambda^n{\cal A}_{\bar{\iota} n}{}^m\p_-X^{\bar{\iota}}\right)+\alpha^{\bar{\iota}}\rho_-^j\bar{\lambda}^n{\cal F}_{\bar{\iota}j n}{}^m\lambda_m
$$
which is of the form seen in  section 3.1.2. This superfield formalism may also be used to describe the ghost fields in the twistor string theory by introducing
$$
B_{--}=\tilde{\beta}_{--}+\theta\tilde{b}_{--}\;,	\qquad	C^-=\tilde{c}^-+\theta\tilde{\gamma}^-\;,	\qquad	B_{++}=b_{++}+\theta f_{++}\;,	\qquad	C^+=c^++\theta g^+\;,
$$
where $f_{++}$ and $g^+$ are arbitrary fields which drop out when we do the $\theta$ integration. The ghost Lagrangian is then
\begin{eqnarray}
{\cal L}_{gh}&=&\int\rd\theta \left(B_{--}\p_+C^-+B_{++}D_-C^+\right)\nonumber\\
&=&b_{++}\p_-c^++\tilde{b}_{--}\p_+\tilde{c}^-+\tilde{\beta}_{--}\p_+\tilde{\gamma}^-\;.
\end{eqnarray}
Note that $B_{++}$ and $B_{--}$ have opposite statistics. A superfield for the $(\bar{\psi}^a_+,\psi^a)$ fields can be constructed along the same lines as the $B_{++}$ and $C^+$ ghost superfields. It is also useful to introduce the stress tensor $S_{\alpha\beta}$ with components
$$
S_{++}=T_{++}\;,	\qquad	S_{--}=T_{--}+\theta G_{--}\;.
$$
The close connection between the topological and conventional Heterotic strings becomes more transparent in this superspace notation and many of the ideas that are useful in the latter may also be applied to the former.

\subsection{Super-Moduli Space for the Topological String}

The previous considerations for the bosonic string are now extended to the twistor string making use of the superfield formalism outlined in the preceding section. The basic aspects of the discussion share many similarities with corresponding considerations in superstring theory and the interested reader is referred to the references \cite{Friedan:1985ge,Verlinde:1988tx,Verlinde:1987sd} for further details. More recently, many of the outstanding issues of superstring perturbation theory have been reconsidered in \cite{Witten:2012bh,Witten:2012ga,Witten:2012bg}.

In the conventional bosonic string theory described above section 4.1 it is necessary to integrate over the moduli $m^K$ associated with deformations of the world-sheet metric $h_{\alpha\beta}$ that cannot be removed by  diffeomorphism or Weyl transformation. In the topological theory it is also necessary to integrate over the moduli $\hat{m}^k$ associated with the gauge fixing of the $Q$-partner of the metric $\chi_{\alpha\beta}$. A reference metric and and its $Q$-partner can be chosen, up to an overall Weyl transformation, as
$$
h_{\alpha\beta}=e^{\Omega}\eta_{\alpha\beta}\;,	\qquad	\chi_{\alpha\beta}=\Omega\nu_{\alpha\beta}\;.
$$
The conformal invariance of the quantum theory then means that $\Omega$ can effectively be set to zero by a Weyl transformation. By analogy with the Heterotic superstring, the path integral expression for a correlation function can be written in terms of an integral over the moduli space of a super-Riemann surface \cite{Verlinde:1988tx} (see \cite{Witten:2012bh,Witten:2012ga,Witten:2012bg} for a recent encyclopaedic treatment in the context of the superstring). It is likely that any attempt to calculate higher loop twistor string amplitudes will be greatly facilitated by using the super manifold formalism. It will be useful in considering the half-twisted model to choose a complex structure on the moduli space and to write the $6g-6+2n$ moduli as $m^K=(m^k,\bar{m}^k)$. The super-moduli space $s{\cal M}_{g,n}$ of an $n$-pointed, genus $g$ super-Riemann surface is a complex manifold of dimension $(6g-6+2n|3g-3+n)$ and is parameterised by the coordinates $(m^k,\bar{m}^k|\hat{m}^k)$. It is helpful to think of $s{\cal M}_{g,n}$ as the space of the super-metrics $(h_{\alpha\beta},\chi_{\alpha\beta})$ up to diffeomorphism and Weyl transformations.

Much of the following analysis could be applied without explicitly importing the language of superfields outlined above. Instead one could consider (0,p)-forms over ${\cal M}_{g,n}$ instead of functions on $s{\cal M}_{g,n}$. In this language the supercoordinates are replaced by a basis of forms in the cotangent space
$$
(\sigma^+,\sigma^-,\theta)\rightarrow (\sigma^+,\sigma^-,\rd\sigma^-)\;,	\qquad	(m^k,\bar{m}_k,\hat{m}^k)\rightarrow (m^k,\bar{m}_k,\rd\bar{m}^k)\;.
$$
The different approaches, writing correlation functions as superfunctions on $s{\cal M}_{g,n}$ or as (0,p)-forms on $\bigwedge^{(0,p)}{\cal M}_{g,n}$, have their own respective advantages. The latter approach is closest to the usual description of correlation functions of the A-model \cite{Hori:2003ic}, whilst the superfield approach allows one to import much of the machinery from conventional ten-dimensional superstring theory.

The starting point is the BRST invariant action $S[\Phi,B]$ written in terms of the matter and ghost superfields $\Phi$ and $B$, respectively. The generalisation to include the additional world-sheet fields is straightforward. Following \cite{Witten:2012bh} and the considerations above for the bosonic case, the BRST symmetry is generalised to one which has a non-trivial action on the pair $(h_{\alpha\beta},\chi_{\alpha\beta})$ so that
$$
\hat{\delta} h_{\alpha\beta}=\mu_{\alpha\beta}\;,	\qquad	\hat{\delta} \mu_{\alpha\beta}=0\;,	\qquad	\hat{\delta} b_{\alpha\beta}=T_{\alpha\beta}
$$
$$
\hat{\delta} \chi_{++}=\hat{\mu}_{++}\;,	\qquad	\hat{\delta} \hat{\mu}_{++}=0\;,	\qquad	\hat{\delta} \tilde{\beta}_{--}=G_{--}
$$
where $T_{\alpha\beta}$ and $G_{--}$ include contributions from ghost \emph{and} matter sectors and $\mu_{\alpha\beta}$ and $\hat{\mu}_{++}$ are changes in $h_{\alpha\beta}$ and $\chi_{++}$ induced by changes in moduli. The action $S[\Phi,B]$ is not be invariant under this generalised $\hat{\delta}$ transformation and instead
$$
\hat{\delta} S=\int_{\Sigma}\rd^2\sigma\left(\mu_{--}T_{++}+\mu_{++}T_{--}+\hat{\mu}_{++}G_{--}\right)\;.
$$
As in the bosonic case, a $\hat{\delta}$-invariant action may be found
$$
\widehat{S}=S+\int_{\Sigma}\rd^2\sigma\left(\mu_{--}b_{++}+\mu_{++}\tilde{b}_{--}+\hat{\mu}_{++}\tilde{\beta}_{--}\right)\;,
$$
which can be expanded out in terms of the $(6g-6+2n|3g-3+n)$ moduli $(m^k,\bar{m}^k|\hat{m}^k)$ of an $n$-pointed super Riemann surface to give
$$
\widehat{S}=S-\int_{\Sigma}\rd^2\sigma\left(\sum_{k=1}^{3g-3+n}\mu_{--,k}b_{++}\rd m^k+\sum_{k=1}^{3g-3+n}\tilde{\mu}_{++k}\tilde{b}_{--}\rd\bar{m}^k+\sum_{k=1}^{3g-3+n}\hat{\mu}_{++k}\tilde{\beta}_{--}\rd\hat{m}^k\right)\;,
$$
so that
\begin{eqnarray}
e^{-\widehat{S}}&=&e^{-S}\prod_{k=1}^{3g-3+n}e^{\langle\mu_k,b\rangle\rd m^k}e^{\langle\tilde{\mu}_k,\tilde{b}\rangle\rd\bar{m}^k}\prod_{k=1}^{3g-3+n}e^{\langle\hat{\mu},\tilde{\beta}\rangle\rd\hat{m}^k}
\nonumber\\
&=&e^{-S}\prod_{k=1}^{3g-3+n}\prod_{k=1}^{3g-3+n}\rd m^k\rd\bar{m}^k\rd\hat{m}^k\delta\left(\langle\mu_k,b\rangle\right)\delta(\langle\tilde{\mu}_k,\tilde{b}\rangle)\delta(\langle\hat{\mu}_k,\tilde{\beta}\rangle)\;,
\end{eqnarray}
where, as in the bosonic case, the fact that $\langle\tilde{\mu}_k,\tilde{b}\rangle$, $\langle\mu_k,b\rangle$, and $\rd\hat{m}^k$ are fermionic has been used. This can be written in terms of superfields defined above. The genus $g$ amplitude for a topological string with operator insertions ${\cal O}_i$ at punctures $\sigma_i$ on $\Sigma$ may be written as
$$
\langle{\cal O}_1...{\cal O}_n\rangle_g=\int_{s{\cal M}_{g,n}}\rd M_k\left\langle\prod_{k=1}^{(6g-6+2n|3g-3+n)}\delta\left(\langle{\cal X}_k,B\rangle\right)\prod_{i=1}^n{\cal O}_i(\sigma_i)\right\rangle\;,
$$
where the supermoduli can be amalgamated into $M^k=(m^k,\bar{m}^k|\hat{m}^k)$ and ${\cal X}_k=(\mu_{++k},\tilde{\mu}_{--k}|\hat{\mu}_{--k})$. This result, whilst highlighting the supergeometry of the moduli space $s{\cal M}_{g,n}$, is perhaps not the most useful for calculating low genus correlation functions. Following the usual arguments in superstring theory, it is instructive to expand $\chi_{\alpha\beta}$ in terms of a particular basis of the fermionic moduli $\hat{m}^k$
$$
\chi_{\alpha\beta}=\sum_{k=1}^{3g-3+n}\chi_{\alpha\beta}^k\hat{m}_k\;.
$$
Note that the $\chi_{\alpha\beta}^k$ are bosonic. Since the $\chi_{++}$ appear linearly in the action (\ref{action1}), the contribution to the action from terms involving $\chi_{++}$ is
$$
S_{\chi}=\sum_{k=1}^{3g-3+n}\hat{m}_k\int_{\Sigma}\rd^2\sigma\sqrt{h}\,\chi_{++}^kG_{--}\;.
$$
Under the extended BRST transformation, the BRST operator acts as an exterior derivative on $s{\cal M}_{g,n}$ \cite{Witten:2012bh} so that $\hat{\delta}\hat{m}_k=\rd\hat{m}_k$, and so
$$
\hat{\delta}\chi_{\alpha\beta}=\sum_{k=1}^{3g-3+n}\chi_{\alpha\beta}^k\rd\hat{m}_k\;.
$$
An action that is invariant under the extended BRST transformation is
$$
\widehat{S}_{\chi}=\sum_{k=1}^{3g-3+n}\left(\hat{m}_k\int_{\Sigma}\rd^2\sigma\sqrt{h}\,\chi_{++}^kG_{--}+\rd\hat{m}_k\int_{\Sigma}\rd^2\sigma\sqrt{h}\,\chi_{++}^k\tilde{\beta}_{--}\right)\;,
$$
so
$$
e^{-\widehat{S}_{\chi}}=\prod_{k=1}^{3g-3+n}e^{\hat{m}_k\langle\chi^k,G\rangle}e^{\rd\hat{m}_k\langle\chi^k,\tilde{\beta}\rangle}=\prod_{k=1}^{3g-3+n}\rd\hat{m}_k\,e^{\hat{m}_k\langle\chi^k,G\rangle}\langle\chi^k,\tilde{\beta}\rangle\;.
$$
Performing the integral over the $\hat{m}_k$ gives
$$
\int e^{-\widehat{S}_{\chi}}=\prod_{k=1}^{3g-3+n}\delta(\langle\chi^k,G\rangle)\langle\chi^k,\tilde{\beta}\rangle
$$
so that the correlation function for the topological string can be written as an integral over the moduli space of an $n$-pointed genus $g$ Riemann surface, not the moduli space of a super Riemann surface \cite{Verlinde:1988tx}
$$
\langle{\cal O}_1...{\cal O}_n\rangle_g=\int_{{\cal M}_{g,n}}\prod_{K=1}^{6g-6+2n}\rd m^K\prod_{k=1}^{3g-3+n}\,\int {\cal D}(X,b,c)\,\Psi_k\,\langle\mu_K,b\rangle\;e^{-S}{\cal O}_1(\sigma_1)...{\cal O}_n(\sigma_n)\;,
$$
where
$$
\Psi_k=\delta(\langle\chi^k,G\rangle)\langle\chi^k,\tilde{\beta}\rangle\;.
$$
A common choice of basis for the moduli $\hat{m}^k$ leads to a simple interpretation of the insertions $\Psi_k$. Choosing $\chi_{++}^k=\delta^2(\sigma^{\alpha}-\sigma^{\alpha}_k)$, which fixes the insertion to be at at the point $\sigma_k$ on $\Sigma$ gives, in this basis, $\langle\chi^k,G\rangle=G_{--}(\sigma_k)$ and $\langle\chi^k,\tilde{\beta}\rangle=\tilde{\beta}_{--}(\sigma_a)$. The $\Psi_k$ is then recognised as a topological analogue of the more familiar picture changing operator \cite{Friedan:1985ge,Verlinde:1988tx}
$$
\Psi_k\rightarrow Y[\sigma_k]=\delta(\tilde{\beta}(\sigma_k))G_{--}(\sigma_k)\;,
$$
and this terminology shall be freely borrowed from superstring theory. The correlation function then reduces to the familiar form of an integral over moduli space ${\cal M}_{g,n}$ with picture changing operator insertions at a finite number of points $\sigma_k$
$$
\langle{\cal O}_1...{\cal O}_n\rangle_g=\int_{{\cal M}_{g,n}}\prod_{K=1}^{6g-6+2n}\rd m^K\prod_{k=1}^{3g-3+3}\,\int {\cal D}(X,b,c)\,Y[\sigma_k]\langle\mu_K,b\rangle\;e^{-S}{\cal O}_1(\sigma_1)...{\cal O}_n(\sigma_n)\;.
$$
This expression may be better understood by recalling the role that pictures play in superstring theory. The $Q$-ghosts have a first order action $\tilde{\beta}_{--}\p_+\tilde{\gamma}^-$ which means there is no lower bound on the energy these fields can have. The potential catastrophe of having bosonic fields with unbounded lower energy is not of concern as there are no interaction terms between these $Q$-ghost fields and other fields in the theory. As such, the $Q$-ghost vacuum is not unique and can be labelled by the picture number $q$ \cite{Verlinde:1990ku}. The picture changing operator is a map from one Bose vacuum to another, each labelled by a different picture number $q$ which give rise to different representations for the same physical state. Examples of vertex operators in the 0 and +1 pictures will be discussed in the following section. An important observation is that, at tree level, non-vanishing scattering amplitudes will have vertex insertions of net picture number +3: a net contribution of +3 from the vertex operators is needed to balance the $Q$-ghost charge of the vacuum for a non-zero correlation function. Tree-level scattering amplitudes will be of primary interest and so only amplitudes constructed from three +1 picture vertices and $n-3$ vertex operators in the 0 picture will be explicitly discussed.

\subsection{Physical Observables}

$n$-particle scattering amplitudes are given by $n$-punctured Riemann surfaces with vertex operators $V$ corresponding to asymptotic physical states inserted at the punctures. In order for this article to be reasonably self-contained, the form that such insertions can take is reviewed in this section. The half-twisted topological gravity modifies the standard arguments in a minor way. Further details regarding vertex operators in half-twisted sigma models or vertex operators in topological gravity may be found in \cite{Mason:2007zv} and \cite{Witten:1989ig} respectively. The vertex operators must preserve the symmetries of the theory. One simple way to construct a diffeomorphsim-invariant operator is to pick a weight $(1,1)$ $\widetilde{Q}_-$-invariant operator ${\cal O}^{(1,1)}$ and integrate it over the world-sheet. Another possibility is to take a weight $(1,0)$ operator ${\cal O}^{(1,0)}$ and attach a puncture operator ${\cal P}$ to it, where ${\cal P}=c^+\tilde{c}^-\delta(\tilde{\gamma}^-)$. The BRST-invariant operators of interest are then of the form
\begin{equation}\label{vertex}
 V={\cal P} \,{\cal O}^{(1,0)}\;,	\qquad	V=\int_{\Sigma}\rd^2\sigma\,{\cal O}^{(1,1)}\;.
\end{equation}
The puncture operator fixes a point on the super-Riemann surface. The ghost insertions in ${\cal P}$ ensure that the diffeomorphisms and $\widetilde{Q}_-$-transformations vanish at the punctures so that the symmetry group is simply given by those diffeomorphisms that preserve the location of the puncture.

Following \cite{Witten:2005px}, consider an operator with dimension $(n,m)=(n,0)$ for non-trivial $Q$-cohomology. The dimensions for the fields $\{X^i,X^{\bar{\iota}},\alpha^{\bar{\iota}},\rho^j_-\}$ are $\{(0,0),(0,0),(0,0),(0,1)\}$. Also, $\partial_-$ has dimension $(0,1)$. Since there are no objects of negative dimension in the theory, we require that observables are only built from the set $\{X^i,X^{\bar{\iota}},\alpha^{\bar{\iota}},\partial_+^kX^i,\partial_+^kX^{\bar{\iota}}\}$ where $k\in\Z$. We neglect the possibility of using $\nabla_+\alpha^{\bar{\iota}}$ and its higher derivative variants as this will vanish on-shell. Thus observables can be expanded in powers of the fermionic field $\alpha^{\bar{\iota}}$ to give
$$
{\cal O}={\cal O}(X,\partial X)+{\cal O}_{\bar{\iota}}(X,\partial X)\alpha^{\bar{\iota}}+\frac{1}{2}{\cal O}_{\bar{\iota}_1\bar{\iota}_2}(X,\partial X)\alpha^{\bar{\iota}_2}\alpha^{\bar{\iota}_2}+\frac{1}{6}{\cal O}_{\bar{\iota}_1\bar{\iota}_2\bar{\iota}_3}(X,\partial X)\alpha^{\bar{\iota}_2}\alpha^{\bar{\iota}_2}\alpha^{\bar{\iota}_3}
$$
$Q$-cohomology is isomorphic to the de-Rahm cohomology of $\C\P^3$. This relationship with de-Rahm cohomology may be seen as follows. Consider a (0,p)-form on $\C\P^3$, where
$$
A=\frac{1}{p!}A_{\bar{\iota}_1\bar{\iota}_2...\bar{\iota}_p}(X)\rd X^{\bar{\iota}_1}\wedge \rd X^{\bar{\iota}_2}\wedge...\wedge\rd X^{\bar{\iota}_p}\;,
$$
and consider the map $(X^I,\rd X^{\bar{\iota}})\rightarrow (X^I,\alpha^{\bar{\iota}})$, then an observable
$$
\mathcal{O}_A=A_{\bar{\iota}_1\bar{\iota}_2...\bar{\iota}_p}(X)\alpha^{\bar{\iota}_1}\alpha^{\bar{\iota}_2}...\alpha^{\bar{\iota}_p}
$$
can be associated to each (0,p)-form on $\C\P^3$. The action of the $Q$ operator on the observable is
$$
\delta_Q\mathcal{O}_A=\{Q,\mathcal{O}_A\}=p\, \alpha^{\bar{\iota}_{p+1}} \partial_{\bar{\iota}_{p+1}}A_{\bar{\iota}_1\bar{\iota}_2...\bar{\iota}_p}(X)\alpha^{\bar{\iota}_1}\alpha^{\bar{\iota}_2}...\alpha^{\bar{\iota}_p}\sim \mathcal{O}_{\bar{\partial}A}
$$
Thus observables in the cohomology of $Q$ are in one-to-one correspondence with elements of the de-Rahm cohomology. A hierarchy of such observables may then be constructed \cite{Mason:2007zv,Witten:2005px,Witten:1991zz}
$$
\bar{\partial}\mathcal{O}^{(n,0)}_A=\{Q,\mathcal{O}_A^{(n,1)}\}\;,	
$$
where $0\leq n\leq 3$. In fact, it is straightforward to see that $Q$ cohomology should be associated to de-Rahm cohomology under this map as
$$
Q=\alpha^{\bar{\iota}}\frac{\delta}{\delta X^{\bar{\iota}}}\rightarrow\rd X^{\bar{\iota}}\frac{\partial}{\partial X^{\bar{\iota}}}=\bar{\partial}\;,
$$
thus $Q^2=0$ directly corresponds to $\bar{\partial}^2=0$. This is one reason why it was useful to consider this nilpotent formulation of topological gravity rather than that given by (\ref{Qsquared}).

It is worth being a little more concrete and apply the above arguments to the closed twistor string theory. Such vertex operators have also been considered in \cite{Mason:2007zv}. Since the Penrose transform relates elements of $H^{(0,1)}(\C\P^3;{\cal O}(-2h-2))$ to physical fields, the vertex operators will be constructed from the following holomorphic (0,1)-forms on $\C\P^{3|4}$
\begin{equation}\label{obs}
{\cal A}_{\bar{\iota}}\;,	\qquad	g_I=(B_{\bar{\iota}j},b_{\bar{\iota}a})\;,	\qquad	f^I=(J^j{}_{\bar{\iota}},j^a{}_{\bar{\iota}})\;,
\end{equation}
The vertex operators in the +1 picture take the form $V={\cal P}\,{\cal O}^{(1,0)}$, where ${\cal P}=c^+\tilde{c}^-\delta(\tilde{\gamma}^-)$ is the puncture operator and
$$
{\cal O}^{(1,0)}_{\cal A}={\cal A}_{\bar{\iota}m}{}^n\alpha^{\bar{\iota}}\bar{\lambda}^m\Gamma_+\lambda_n\;, \qquad	{\cal O}^{(1,0)}_B=B_{\bar{\iota}j}\alpha^{\bar{\iota}}\partial_+X^j\;,	\qquad	{\cal O}^{(1,0)}_J=g_{\bar{\iota}j}J^j{}_{\bar{k}}\alpha^{\bar{k}}\partial_+X^{\bar{\iota}}\;,
$$
$$
{\cal O}^{(1,0)}_b=b_{\bar{\iota}a}\alpha^{\bar{\iota}}\partial_+\psi^a\;,	\qquad	{\cal O}^{(1,0)}_j=j^a{}_{\bar{\iota}}\alpha^{\bar{\iota}}\bar{\psi}_{+a}\;.
$$
The 0 picture vertex operators take the form
$$
V=\oint_{\Sigma}\rd^2\sigma\,{\cal O}^{(1,1)}\;.
$$
Given the form of the picture changing operator found in the last section, it is clear that these 0 picture operators are given by ${\cal O}^{(1,1)}=\{\widetilde{Q}_-,{\cal O}^{(1,0)}\}$
$$
{\cal O}^{(1,1)}_{\cal A}={\cal A}_{\bar{\iota}m}{}^n\partial_-X^{\bar{\iota}}\bar{\lambda}^m\Gamma_+\lambda_n\;, \qquad	{\cal O}^{(1,1)}_B=B_{\bar{\iota}j}\partial_-X^{\bar{\iota}}\partial_+X^j\;,	\qquad	{\cal O}^{(1,1)}_J=g_{\bar{\iota}j}J^j{}_{\bar{k}}\partial_-X^{\bar{k}}\partial_+X^{\bar{\iota}}\;.
$$
$$
{\cal O}^{(1,1)}_b=b_{\bar{\iota}a}\partial_-X^{\bar{\iota}}\partial_+\psi^a\;,	\qquad	{\cal O}^{(1,1)}_j=j^a{}_{\bar{\iota}}\partial_-X^{\bar{\iota}}\bar{\psi}_{+a}\;.
$$
where the fact that the restriction of the of $F^{(1,1)}$ to $\Sigma$ vanishes has been used. Only terms that come from the $\widetilde{Q}_-$ variation of the field that appear in the (2,0) sigma model have been explicitly written down. The fact that $\widetilde{Q}_-^2=0$ ensures that the ${\cal O}^{(1,1)}$ are $\widetilde{Q}_-$-invariant. Of particular  interest to twsitor string theory are vertex operators constructed from spinor helicity variables $\{\lambda_{A'},\widetilde{\lambda}_{A}\}$, which encode the on-shell momenta $P_{AA'}=\lambda_{A'}\widetilde{\lambda}_{A}$ of an asymptotic momentum eigenstate in space-time. For example, the twistor representatives for the ${\cal N}=4$ Yang-Mills multiplet gives
\begin{eqnarray}
{\cal A}&=&\int\frac{\rd\xi}{\xi}\delta^2(\lambda_{A'}-\xi\pi_{A'}(\sigma)) e^{-i\xi\omega^A(\sigma)\widetilde{\lambda}_A}\left(A_{(0)}+\xi A_{(-1)a} \psi^a+\frac{1}{2}\xi^2A_{(-2)ab}\psi^a\psi^b\right.\nonumber\\
&&\left.+\frac{1}{6}\xi^3A_{(-3)}^a\varepsilon_{abcd}\psi^b\psi^c\psi^d+\frac{1}{24}\xi^4A_{(-4)}\varepsilon_{abcd}\psi^a\psi^b\psi^c\psi^d\right)
\end{eqnarray}
There are similar expressions for the vertex operators constructed from the superfields $f^I$ and $g_I$ in (\ref{obs}) (see \cite{Dolan:2008gc} for more details).

 \subsection{Yang-Mills Tree Amplitudes}

The above constructions may be applied to reconstruct the known twistor string derivation of the $n$-point gluon tree-level scattering amplitude. As found in section 4.2, the $g$-loop scattering amplitudes are given by
$$
\langle V_1...V_n\rangle_g=\int_{{\cal M}_{g,n}}\prod_{K=1}^{6g-6+2n}\rd m^K\prod_{k=1}^{3g-3}\,\int {\cal D}(X,b,c)\,Y[\sigma_k]\langle\mu_K,b\rangle\;e^{-S}\,V_1(\sigma_1)...V_n(\sigma_n)\;,
$$
At tree level ($g=0$) the world-sheet is a $\C\P^1$ and there are no moduli (so there are no $b_{\alpha\beta}$'s or $\beta_{\alpha\beta}$'s). The expression for the amplitude then becomes
$$
\langle V_1...V_n\rangle_g=\int {\cal D}(X,c,\gamma)\,e^{-S}\,\prod_{i=1}^3V^{(+1)}_i\prod_{i=4}^nV^{(0)}_i(\sigma_i)\;,
$$
where the vertex operators $V^{(p)}$ in the $p$'th picture are
$$
V^{(0)}_i=\int_{\Sigma}\rd^2\sigma_i\,\bar{\lambda}^m\Gamma_+\lambda_n{\cal A}_{\bar{\iota}m}{}^n\p_-X^{\bar{\iota}}\;,	\qquad	V^{(+1)}_i(\sigma_i)=c_i\tilde{c}_i\delta(\tilde{\gamma}_i)\bar{\lambda}^m\Gamma_+\lambda_n{\cal A}_{\bar{\iota}m}{}^n\alpha^{\bar{\iota}}\;.
$$
These vertex opertaors are related to each other by the picture changing operator $Y[\sigma_i]$. The total $Q$-ghost anomaly must vanish, requiring 3 picture +1 vertex operators and the remaining $n-3$ vertex operators to be in the 0 picture. The scattering amplitude of interest is given by the path integral
$$
\langle V_1...V_n\rangle_0=\int {\cal D}(X,\alpha,\lambda, c ,\gamma) \;e^{-S}\,\prod_{i=1}^3c_i\tilde{c}_i\delta(\tilde{\gamma}_i)\bar{\lambda}^m\Gamma_+\lambda_n{\cal A}_{\bar{\iota}m}{}^n\alpha^{\bar{\iota}}\prod_{i=4}^n \int_{\Sigma}\rd^2\sigma_i\,\bar{\lambda}^m\Gamma_+\lambda_n{\cal A}_{\bar{\iota}m}{}^n\p_-X^{\bar{\iota}}\;.
$$
In order to have a meaningful correlation function, it is necessary that all of the anomaly conditions, discussed in section 3.3, are satisfied. The result of the $\psi$ functional integration will be zero unless $d=k-1$ where $d$ is the degree of the holomorphic map to twistor space and $k+2$ is the number of negative helicity gluons in the amplitude. Further details of this scattering amplitude calculation are given in Appendix B. The result is the scattering amplitude may then be written in the familiar form given in \cite{Roiban:2004yf}.

\section{A Local First Order Description}

In \cite{Witten:2005px} it was argued that the (2,0) model on a patch $U\subset\C\P^3$ could be described in terms of a first order system on $U$. In \cite{Mason:2007zv} this was used to suggest a relationship between the heterotic twistor string and a closed version of the Berkovits string \cite{Berkovits:2004hg}. With the role of the $Q$-ghost sectors better understood, this correspondence will be reconsidered in this section.

In \cite{Witten:2005px}, Witten argued that the information relating to the target space metric was Q-exact, in other words it should not affect the cohomology and will therefore not be observable. One is then free to choose a patch $U$ of $\C\P^{3}$ with a flat metric so that the gauge-fixed sigma model simplifies to
\begin{eqnarray}
S_U&=&\int_{\Sigma}\rd^2 \sigma\; \left(\delta_{\bar{\iota}j}\partial_- X^j\partial_+X^{\bar{\iota}}-i\bar{\psi}^a_+\p_-\psi_a-i\delta_{\bar{\iota}j}\rho_-^j\partial_+\alpha^{\bar{\iota}}-i\bar{\lambda}^m\Gamma_+{\cal D}_-\lambda_m\right.\nonumber\\
&&\left.+b_{++}\p_-c^++\tilde{b}_{--}\p_+\tilde{c}^-+\tilde{\beta}_{--}\p_+\tilde{\gamma}^-\right)\;.
\end{eqnarray}
It is then useful to define the world-sheet (1,0) forms $P_{+j}:=\delta_{\bar{\iota} j}\p_+X^{\bar{\iota}}$ and use them to rewrite the sigma model in first order form
$$
S_U=\int_{\Sigma}\rd^2 \sigma\; \left(P_{+i}\partial_-X^i-i\bar{\psi}^a_+\p_-\psi_a-i\bar{\lambda}^m\Gamma_+{\cal D}_-\lambda_m+b_{++}\p_-c^++\tilde{b}_{--}\p_+\tilde{c}^-+\tilde{\beta}_{--}\p_+\tilde{\gamma}^-\right)\;.
$$
The ($\rho^j_-,\alpha^{\bar{\iota}}$) sector has also been removed. The idea is that a zero contribution to the conformal anomaly has been removed (the ($\rho,\alpha$) system contributed (0,-D) and the $\p_+ X^{\bar{\iota}}$ contributed (0,D) so there is no net change in the central charge) and the removed sectors did not contribute to observables of the model (at least not perturbatively) \cite{Witten:2005px}. It is simple to rewrite the central charge count for the first order twistor string and check that the conformal anomaly is cancelled for $N=28$. The gauge-fixed action can be expressed in terms of homogenous coordinates on the $\C\P^{3|4}$ as discussed in section 3.1, where we now combine the bosonic and fermonic components of the homogenous coordinates of the $\C\P^{3|4}$ into the single supercoordinate $Z^I$
$$
S_U=\int_{\Sigma}\rd^2 \sigma\; \left(Y_I\bar{D}Z^{I}-i\bar{\lambda}^m\bar{\cal D}\lambda_m+b\,\bar{\p}c+\tilde{b}\,\p\tilde{c}+\tilde{\beta}\,\p\tilde{\gamma}\right)\;,
$$
where complex world-sheet coordinates have been used (as is conventional when describing this system\footnote{Note that the choice of invariant metric is now taken in conformal gauge $h_{\alpha\beta}=e^{\varphi}\delta_{\alpha\beta}$.}) and $\bar{D}Z^I=\bar{\p}Z^I+\bar{A}Z^I$ is a $GL(1;\C)$ covariant derivative. Using the $\bar{A}$ equation of motion $Y_0=-Y_iZ^i/Z^0$, one can relate the given fields \cite{Mason:2007zv}: $X^i=Z^i/Z^0$, $\psi^a=Z^a/Z^0$, $P_{+i}=Z^0Y_{+i}$, and $\bar{\psi}_{+a}=Z^0Y_{+a}$. In the above action $Z^I=(Z^0,Z^i,Z^a)$ are homogenous coordinates on $\C\P^{3|4}$. Superficially, this looks like the Berkovits string \cite{Berkovits:2004hg}; however, topological gravity has been gauge-fixed which leads to the introduction of a ($\tilde{\beta},\tilde{\gamma}$) ghost system not present in the usual gauge-fixed Berkovits model. There is also the obvious fact that this model is a closed string with different fields in the holomorphic and anti-holomorphic sectors, not an open one. The field content, including the contribution to central charges is given in the table below. The projective symmetry is gauge-fixed ($\bar{A}=0$) and the corresponding ghosts $u$ and $v$ are introduced via the standard Fadeev-Popov procedure.

\begin{table}[htdp]
\caption{Closed Twistor String Field Content}
\begin{center}
\begin{tabular}{|c c c c | c c c c|}
\hline
&  Left-Movers & & &  & Right-Movers & & \\
\hline\hline
Field & Central Charge & Spin & Statistics & Field & Central Charge & Spin & Statistics \\
\hline
&&&&&&&\\
$Y_I,Z^I$		&	$(0,0)$ & 1,0 & \text{Bose} & 	&&& 	\\
$b,c$		&	$(-26,0)$ & 2,1 & \text{Fermi}  & $\tilde{b},\tilde{c}$		&	$(0,-26)$ & 2,1 & \text{Fermi}\\
$\lambda^m$	&	$(+28,0)$ & $\frac{1}{2}$ & \text{Fermi}  & $\tilde{\beta},\tilde{\gamma}$	&	$(0,+26)$	& 2, 0 & \text{Bose}\\
$u$, $v$	&	$(-2,0)$ & 1,0 & \text{Fermi} & &&& \\
&&&&&&&\\
\hline
\end{tabular}
\end{center}
\label{default}
\end{table}%

A closed model of this form has been used as the basis for pioneering investigations into ${\cal N}=8$ supergravity and twistor string theory \cite{Adamo:2012xe,Adamo:2012nn} although, until now the role of topological gravity and the precise details of the conformal anomaly cancellation have been unclear. The calculation of scattering amplitudes should be straightforward, with the $Q$-ghosts now playing a role and vertex operators involving the pull-back (0,1)-forms (Dolbeault twistor representatives) ${\cal A}(Z)$ and $f_I(Z)\rd Z^I$ and the vector field $g^I\frac{\partial}{\partial Z^I}$ to the world-sheet to give the closed string vertex operators \cite{Berkovits:2004jj}. The integrated vertex operators are now
$$
V_{\cal A}=\int_{\Sigma}{\cal A}_m{}^n\wedge\bar{\lambda}_n\partial \lambda^m\;,	\qquad	V_f=\int_{\Sigma}f_I\wedge\partial Z^I\;,	\qquad	V_g=\int_{\Sigma}g^I\wedge Y_I\;,	\qquad
$$
It is interesting to wonder whether there is a way to introduce topological gravity into a non-chiral closed string on which one might impose real boundary conditions so that the observables of the theory take values in a real slice of twistor space corresponding to a space-time of split-signature (see Appendix A for a very brief introduction to twistor geometry including issues of space-time signature). This would yield an open string, similar to the Berkovits string, but arising from a gauge fixing of topological, not conventional, world-sheet gravity. A suitable candidate for such a string theory with topological gravity is not immediately obvious, nonetheless the gauge-fixed action
\begin{equation}\label{topBerkovits}
S_U=\int_{\Sigma}\left(Y_I\bar{\p}Z^{I}+\tilde{Y}_I\p\tilde{Z}^{I}-i\lambda^m\bar{\cal D}\lambda_m-i\tilde{\lambda}^m{\cal D}\tilde{\lambda}_m+b\bar{\p}c+\tilde{b}\p\tilde{c}+\beta\bar{\p}\gamma+\tilde{\beta}\p\tilde{\gamma}+u\bar{\p}v+\tilde{u}\p\tilde{v}\right)
\end{equation}
could be expected to be the result of gauge-fixing such a system. The $(u,v)$ and $(\tilde{u},\tilde{v})$ fields are ghosts introduced to gauge fix the projective $GL(1;\C)$ symmetry. In order to make sense of such an action as a gauge-fixed theory, one would need to understand how to couple topological, as opposed to conventional, gravity to the Berkovits string. Alternatively, one might take the action (\ref{topBerkovits}) as the starting point in defining the string theory. The crucial point is that in this model model the central charge of the gauge sector is required to be $+2$, a figure that may be more easy to reconcile with the anomaly cancellation results of \cite{Romer:1985yg}.

\section{Discussion}

The nature of world-sheet gravity for the closed twistor string has been clarified. In particular, the conformal anomaly is shown to cancel on both left and right moving modes. The inclusion of the $Q$-partner for the world-sheet metric allows a natural interpretation of twistor string scattering amplitudes in terms of an integral over the moduli space of super-Riemann surfaces along the lines of the standard treatment of the conventional critical superstring. The inclusion of topological gravity in the closed twistor string suggests a role for topological gravity in the open twistor string. Though a complete description has not yet been obtained, there is some evidence that such a coupling of the open twistor string to topological gravity will reduce the rank of the gauge group of the theory.

Several questions still remain. In conventional string theory, the dilaton expectation value controls the string coupling via a Fradkin-Tseytlin type interaction term ${\cal L}=\sqrt{-h}\,\varphi \,R_{\Sigma}$, where $R_{\Sigma}$ is the Ricci scalar for world-sheet gravity. It is unclear what candidates there are for the scalar field $\varphi$. It is clear that a scalar on $\C\P^3$ does not correspond to a physical field in the target space via the Penrose transform. Given this, the rationale for twistor string perturbation theory cannot be the standard one. Indeed, in \cite{Berkovits:2004jj} it was argued that the genus expansion comes from the coupling of the B-field in twistor space. This makes sense for a closed string and forces the conformal gravity to be of an exotic form, about which very little is known \cite{Berkovits:2004jj}. It would be good to have a better understanding of this theory from a world-sheet perspective.

The treatment of pictures is reliable if we stay away from the boundary of moduli space, where Riemann surfaces degenerate. Unfortunately, this is the most important region for considering the BCFW recursion relation, as this is where string propagators go on-shell. Witten has recently revisited this issue \cite{Witten:2012bh} and it would be interesting to see to what extent these issues might be resolved by this approach. Most of the considerations presented here generalise straightforwardly to twistor strings for self-dual supergravity \cite{AbouZeid:2006wu}.

\begin{center}
\textbf{Aknowledgements}

I would like to thank Lionel Mason for various conversations. I am supported by Stipendiary Lectureships at Merton and St. Peter's College, Oxford and also at the Mathematical Institute in Oxford.
\end{center}

\begin{appendix}

\section{Twistor Theory}

In this Appendix we briefly review those aspects of twistor theory most relevant to this article. Standard references on twistor theory are \cite{Ward:1990vs,Huggett:1986fs}. Many of the recent advances in calculating scattering amplitudes using twistor theory are summarised in \cite{Adamo:2011pv,ArkaniHamed:2009dn,ArkaniHamed:2010kv}. The application to six-dimensions has also been explored \cite{Mason:2011nw,Saemann:2011nb}.

Super-twistor space is a $\C\P^{3|4}$ with homogenous coordinates $(Z^I,\psi^a)\sim (tZ^I,t\psi^a)$ where $t\in\C/{\{0\}}$. It is convenient to chose decompose the bosonic part of the homogenous twistor into primary and secondary parts $Z^I=(\omega^A,\pi_{A'})$.

Fields on space-time, satisfying  Incidence relation
$$
\omega^A=x^{AA'}\pi_A'\;,	\qquad	\psi^a=\theta^{A'a}\pi_{A'}
$$
relates a point on chiral superspace $(x^{AA'},\theta^{A'a}\pi_{A'})$ to a Riemann sphere in twistor space $S_x=\C\P^1\subset \C\P^{3|4}$.

The direct Penrose transform relates an element of de-Rahm cohomology $f(Z)$ of homogeneity $2h-2$ to a solution of the (linearised) field equations for a massless spin $h$ field in space-time.
\begin{equation}\label{PT}
\Phi_{A'_1...A'_n}(x)=\oint_{S_x}\D\pi\wedge \pi_{A'_1}...\pi_{A'_1}\,f(x\cdot\pi,\pi)\;,
\end{equation}
where $\D\pi=\varepsilon^{AB}\pi_A\rd\pi_B$ and $f(x\cdot\pi,\pi)\in H^{1}_{\bar{\partial}}({\cal O}(2h-2))$. This formula can be understood in terms of the correspondence space $\F$ which may be thought of as a $\C\P^1$ bundle over complexified space-time $\C\M\sim \C^4$ with coordinates $(x^{AA'},[\pi])$.
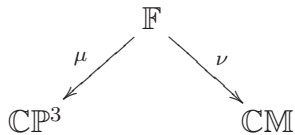
\begin{align*}
\xymatrix{& \F \ar[dl]_\mu \ar[dr]^\nu & \\
\C\P^3 & & \C\M }
\end{align*}
Given a twistor representative $f(Z)$ on $\C\P^3$, we can pull it back to $\F$ to give $\mu^*f$, one then integrates over the fibre of the $\mu$-fibration to get the space-time field giving the relation (\ref{PT}). The indirect transform, which relates an element of $H^{0,1}(S_x;{\cal O}())$ be written as
$$
\Psi^{A_1...A_n}(x)=\oint_{S_x}\D\pi\wedge \frac{\partial^n}{\partial \omega^{A_1}...\omega^{A_n}}\,f(x\cdot\pi,\pi)\;,
$$
Historically, the major obstacle in twistor theory has been introducing non-linear interactions. Twistor string theory an its spin-offs represent a great advance in that respect. For self-dual non-abelian Yang-Mills fields the correspondence is known as the Penrose-Ward correspondence \cite{Ward:1977ta}.

Real space-times $\M\subset\C\M$ with signature $(p,q)$ are given by the fixed points of the map $\tau:x^{AA'}\rightarrow x^{AA'}$, where we think of $x^{AA'}$ as a $2\times 2$ complex matrix. The action of $\tau$ induces an action on twistor space $\T$.

\begin{itemize}
\item \textbf{Split Signature}: $\tau(x^{AA'})=(x^{AA'})^*$ is the standard complex conjugation of the entires of the matrix. The real subspace is given by the \emph{real} $x^{AA'}$. Thus $\M=\R^4\subset\C\M$, which is preserved by $SL(4;\R)=Spin(3,3)$. On twistor space $\tau$ acts by complex conjugation on the components of a twistor
$$
\tau: Z^{\alpha}\rightarrow (Z^{\alpha})^*
$$
so it selects the real slice $\R^4\subset\C^4$. Projective twistor space is then $\R\P^3\subset\C\P^3$.
\item\textbf{Minkowski Signature}: $\tau(x^{AA'})=(x^{AA'})^{\dagger}$ where $\dagger$ is the standard Hermitian conjugation of the matrix. This subspace is preserved by $SU(2,2)=Spin(4,2)$. On twistor space $\tau$ acts by complex conjugation on the components of a twistor
$$
\tau: Z^{\alpha}\rightarrow \bar{Z}_{\alpha}=\Sigma_{\alpha\bar{\beta}}\bar{Z}^{\bar{\beta}}
$$
where $\bar{Z}^{\bar{\alpha}}=(Z^{\alpha})^*$ and $\Sigma_{\alpha\bar{\beta}}$ is the Hermitian metric preserved by the conformal group $SU(2,2)$. This map may be written as $\tau:\bar{\T}\rightarrow\T^*$, thus identifying a twistor with its dual.
\item\textbf{Euclidean Signature}: $\tau(x^{AA'})=\hat{x}^{AA'}$ where $\hat{}$ denotes Hermitian conjugation
$$
\hat{x}=-\sigma_2 x^*\sigma_2
$$
where $\sigma_2$ is a Pauli matrix. The real subspace is preserved by $SL(2;\mathbb{H})=Spin(5,1)$. On twistor space $\tau$ acts by Hermitian conjugation
$$
\tau: Z^{\alpha}\rightarrow \hat{Z}^{\alpha}
$$
where if $Z^{\alpha}=(\omega^A,\pi_{A'})$, then $\hat{Z}^{\alpha}=(\hat{\omega}^A,\hat{\pi}_{A'})$, where $\hat{\omega}^A=(\bar{\omega}^1,-\bar{\omega}^0)$ and similarly for $\hat{\pi}_{A'}$.
\end{itemize}

\noindent It is of interest to understand how the map $\tau$ can be imposed on the twistor string theory, possibly in conjunction with an operation on the world-sheet ${\cal I}(\sigma)=\sigma$, to give a string theory which calculates physical observables in these space-time signatures. To date, the only satisfactory construction has been that of Berkovits \cite{Berkovits:2004hg} in which the string endpoints are restricted to lie in the real slice $\R\P^{3|4}$ and ${\cal I}$ is simply complex conjugation on the world-sheet. This gives an open string theory with end-points on the real slice $\R\P^{3|4}$.

\section{Yang-Mills Tree Amplitudes}

The calculation of Yang-Mills tree amplitudes is sketched in this appendix. Further details may be found, for sample in \cite{Mason:2007zv}. The scattering amplitude of interest is given by the path integral (section 4.4)
$$
\langle V_1...V_n\rangle_0=\int {\cal D}(X, c ,\gamma) \;e^{-S}\,\prod_{i=1}^3c_i\tilde{c}_i\delta(\tilde{\gamma}_i)\bar{\lambda}^m\lambda_n{\cal A}_{\bar{\iota}m}{}^n\alpha^{\bar{\iota}}\prod_{i=4}^n \int_{\Sigma}\rd^2\sigma_i\sqrt{h}\,\bar{\lambda}^m\lambda_n{\cal A}_{\bar{\iota}m}{}^n\p_-X^{\bar{\iota}}\;.
$$
The ghost parts can be evaluated first. The three insertions of $\delta(\tilde{\gamma}_i)$ balance the -3 net ghost charge of the vacuum. Similarly, the three $c_i\tilde{c}_i$ insertions balance the -6 vacuum ghost charge. So that the net effect of the ghost sector is
$$
\langle c(\sigma_1)\tilde{c}(\sigma_1)c(\sigma_2)\tilde{c}(\sigma_2)c(\sigma_3)\tilde{c}(\sigma_3)\rangle=\frac{1}{\text{Vol(GL(2))}}\;,
$$
so that we have
$$
\langle V_1...V_n\rangle_0=\int {\cal D}(X,\alpha,\psi,\lambda) \frac{1}{\text{Vol(GL(2))}}\;e^{-S}\,\prod_{i=1}^3\bar{\lambda}^m\lambda_n{\cal A}_{\bar{\iota}m}{}^n\alpha^{\bar{\iota}}\prod_{i=4}^n \int_{\Sigma}\rd^2\sigma_i\sqrt{h}\,\bar{\lambda}^m\lambda_n{\cal A}_{\bar{\iota}m}{}^n\p_-X^{\bar{\iota}}\;.
$$
The gauge-fixed BRST-invariant action is
\begin{eqnarray}
S&=&\int_{\Sigma}\rd^2 \sigma\;e \,g_{\bar{\iota}j}\left(\partial_- X^j\partial_+X^{\bar{\iota}}-i\rho_-^j\left(\nabla_+\alpha^{\bar{\iota}}+\chi_{++}\partial_-X^{\bar{\iota}}\right)\right)-i\int_{\Sigma}\rd^2 \sigma\;e\left(\bar{\psi}^a_+\p_-\psi_a+\;\bar{\lambda}^m{\cal D}_-\lambda_m\right)\nonumber\\
&&+\int_{\Sigma}\rd^2 \sigma\;e\left(b_{++}\p_-c^++\tilde{b}_{--}\p_+\tilde{c}^-+\tilde{\beta}_{--}\p_+\tilde{\gamma}^-\right)\;.\nonumber
\end{eqnarray}

Following \cite{Mason:2007zv} we deal with the $\lambda$ integration by making the replacement
$$
\int{\cal D}\lambda e^{-S}\;{\cal F}(\lambda)\rightarrow \left[{\cal F}\left(\frac{\delta}{\delta J}\right)\int{\cal D}\lambda \;e^{-S-\oint_{\Sigma}\text{tr}\bar{\lambda} J \lambda}\right]_{J=0}\;.
$$
Since
$$
\int{\cal D}\lambda\;e^{-\oint_{\Sigma}\text{tr}\lambda ({\cal D}_-+J)\lambda}=\text{det}({\cal D}_-+J)\;,
$$
it was argued in \cite{Mason:2007zv} that
$$
\left[\frac{\delta}{\delta J}\text{det}({\cal D}_-+J)\right]_{J=0}= G(\sigma_i,\sigma_j)=\frac{\rd \sigma_i}{\sigma_i-\sigma_j}
$$
is the Green function for the free $\lambda$ system. The functional derivatives will give single trace and multi-trace terms. It is perhaps to be expected that the closed string would have single and multi-trace terms, even at tree level since there is no canonical way to order the vertices. It is thought that the multi-trace terms contribute to conformal supergravity so we neglect them here, although it is interesting that the Yang-Mills vertices seem to contain information about the conformal graviton. The single-trace contributions give (up to unimportant normalisations)
$$
\langle V_1...V_n\rangle_0=\int {\cal D}(X,\alpha,\psi) \frac{1}{\text{Vol(GL(2))}} \;e^{-S}\,\text{tr}
\prod_{i=1}^3\frac{\rd \sigma_i}{\sigma_i-\sigma_{i+1}}{\cal A}_{\bar{\iota}m}{}^n\alpha^{\bar{\iota}}\prod_{i=4}^n \int_{\Sigma}\frac{\rd^2\sigma_i}{\sigma_i-\sigma_{i+1}}
{\cal A}_{\bar{\iota}m}{}^n\p_-X^{\bar{\iota}}
$$
There are no $\rho^j_-$ modes in the observables, so the condition that the $(\alpha,\rho)$ system is anomaly free is
$$
N_{\alpha}=4d+3
$$
The expression for the amplitude has three $\alpha^{\bar{\iota}}$'s so an extra $4d$ of them are required to satisfy the anomaly condition. The origin of such modes was discussed in \cite{Mason:2007zv} and is related to the question of how one chooses a contour on the space of maps over which to integrate. We refer the reader to \cite{Mason:2007zv} for details and, following the example set in \cite{Mason:2007zv}, proceed to integrate over the $\alpha^{\bar{\iota}}$

Given a point $\sigma_i$ on $\Sigma$, the embedding of a degree $d$ curve into supertwistor space is given by
$$
Z^I_i=\sum_{k=0}^dZ_k^I\sigma^k_i
$$
where the coordinates $Z^I=(\omega^A,\pi_{A'})$ are the homogenous coordinates of section 3.1. It is more convenient for us to consider the embedding in terms of these homogenous coordinates $(Z^I,\psi^a)$. For genus zero, the space of degree $d$ maps $X:\Sigma\rightarrow \C\P^3$ is
$$
\text{Maps}(\Sigma,\C\P^3)=\C\P^{4d+3}\;,
$$
with measure
$$
\rd^{4d+3}Z\rd^{4d+4}\psi=\D^{2d+1}\pi\rd^{4d+4}\omega\rd^{2d+2}\psi\;.
$$
We consider the twistor representatives for positive and negative helicity gluons, (0,1)-forms of projective weight 0 and -4
$$
A_{-4}=\int \rd\xi_i\,\xi^3\,\delta^2(\lambda_{iA'}-\xi_i\pi_A'(\sigma_i))e^{-i\xi_i\omega^A(\sigma_i)\widetilde{\lambda}_{iA}}\;,	\qquad	A_0=\int \frac{\rd\xi_i}{\xi_i}\,\delta^2(\lambda_{iA'}-\xi_i\pi_A'(\sigma_i))e^{-i\xi_i\omega^A(\sigma_i)\widetilde{\lambda}_{iA}}\;.
$$
Integrating over the $\omega^A_k$ and $\psi^a_k$ gives the terms
$$
\prod_{k=0}^d\delta^2\left(\sum_{i=1}^n\xi_i\sigma^k_i\widetilde{\lambda}_i^{A}\right)\delta^4\left(\sum_{i=1}^n\xi_i\sigma^k_i\eta_i^{A}\right)\;.
$$
In order to have a meaningful correlation function, we require that the anomaly conditions are satisfied. The result of the $\psi$ integrals will be zero unless $d=n-1$ where $d$ is the degree of the holomorphic map to twistor space and $n$ is the number of negative helicity gluons in the amplitude. The scattering amplitude may then be written in the familiar form \cite{Roiban:2004yf}
$$
\langle V_1...V_n\rangle_0=\sum_{d=1}^{n-3}\int\frac{\rd^{2d+2}\pi_k\rd^n\sigma\rd^n\xi}{\text{Vol(GL(2))}}\prod_{i=1}^n\frac{\delta^2\left(\lambda_{iA'}-\xi_i\pi_{A'}(\sigma_i)\right)}{\xi_i(\sigma_i-\sigma_{i+1})}\prod_{k=0}^d\delta^2\left(\sum_{i=1}^n\xi_i\sigma^k_i\widetilde{\lambda}_i^{A}\right)\delta^4\left(\sum_{i=1}^n\xi_i\sigma^k_i\eta_i^{A}\right)\;.
$$

\end{appendix}

\end{document}